# Dielectric Spectroscopy on Organic Charge-Transfer Salts

## P. Lunkenheimer and A. Loidl

Experimental Physics V, Center for Electronic Correlations and Magnetism, University of Augsburg, 86159 Augsburg, Germany

E-mail: peter.lunkenheimer@physik.uni-augsburg.de

**Abstract**
This Topical Review provides an overview of the dielectric properties of a variety of organic charge-transfer salts, based on both, data reported in literature and our own experimental results. Moreover, we discuss in detail the different processes that can contribute to the dielectric response of these materials. We concentrate on the family of the one-dimensional (TMTTF)$_2$X systems and the two-dimensional BEDT-TTF-based charge-transfer salts, which in recent years have attracted considerable interest due to their often intriguing dielectric properties. We will mainly focus on the occurrence of electronic ferroelectricity in these systems, which also includes examples of multiferroicity.

___

**Contents**

# 1. Introduction

Organic charge-transfer salts exhibit a rich variety of different electronic and magnetic phases, including Mott insulators, unconventional superconductors, charge order (CO), spin- and charge-density waves (CDW), spin liquids, ferroelectrics and even multiferroics. Moreover, different dimensionalities are found in these systems ranging from three- to one-dimensional. Especially for the investigation of the ferroelectric phases, dielectric spectroscopy is inevitable and for various members of this material class intriguing and unexpected dielectric behaviour was found (e.g., [1,2,3,4,5,6,7,8,9,10,11]). Moreover, CO, which is a rather widespread phenomenon in organic charge-transfer salts [12], is usually accompanied by a significant increase of resistivity or even a metal-insulator transition when the formerly delocalized electrons condense into the charge-ordered phase. CO is often accompanied by dielectric anomalies and interesting dielectric properties can result from this phenomenon, e.g., so-called colossal dielectric constants [13,14,15]. Special interest arises from the fact that in some organic charge-transfer salts a CO-driven ferroelectric state was detected (e.g., [1,2,3,4,16]) characterizing them as so-called electronic ferroelectrics. In contrast to the off-centre displacement of ions in canonical ferroelectrics, in this class of materials the *electronic* degrees of freedom generate polar order. This exotic phenomenon has attracted considerable interest and, when combined with magnetic ordering, can give rise to new classes of multiferroic materials. Multiferroicity, especially the simultaneous occurrence of dipolar and magnetic order, is a relatively rare phenomenon [17]. Due to its promising applications in future electronics, the search for new multiferroics and the investigation of the underlying microscopic mechanisms is one of the most active fields in current material science [18,19,20]. Indeed organic charge-transfer salts were considered as good candidates for multiferroics with a ferroelectric state that is driven by CO [21].

In this Topical Review, we will mainly concentrate on two families of organic charge-transfer salts, the quasi-one-dimensional systems (TMTTF)$_2X$ (TMTTF: tetramethyltetrathiofulvalene) and the two-dimensional (ET)$_2X$ systems involving layers of bis(ethylenedithio)-tetrathiafulvalene molecules (often abbreviated as BEDT-TTF or ET). Members of the first group of materials belonged to the first examples of electronic ferroelectricity [1,2]. They feature uniform stacks of nearly planar organic TMTTF molecules, forming chainlike structures with rather strong dimerization of neighbouring molecules. This dimerization, combined with charge ordering, gives rise to the ferroelectric state [21]. Interestingly, only in part of the ferroelectric one-dimensional charge-transfer salts ferroelectricity arises from the electronic degrees of freedom. As an example, tetrathiafulvalene-p-bromanil (TTF-BA) will be briefly treated, where mixed anion-cation stacks exist and the ferroelectricity is mainly driven by ionic displacements [5]. Especially interesting is the case of tetrathiafulvalene-p-chloranil (TTF-CA) [6]. While its structure also comprises mixed stacks of the two molecular species, nevertheless it seems that electronic degrees of freedom dominate over the ionic ones in this system.

The dielectric properties of the two-dimensional (ET)$_2X$ systems have attracted considerable interest in recent years. Their structure includes layers of ET molecules, which can be arranged in a large variety of different patterns (indicated by a Greek letter preceding the (ET)$_2$ part of the formula) which, besides of the anion layers, strongly affects their physical properties. For example, recently multiferroicity was found in $\kappa$-(ET)$_2$Cu[N(CN)$_2$]Cl [7]. Moreover, in [7] a new electric-dipole driven mechanism of multiferroicity was proposed to explain the unexpected fact that the magnetic and polar order appear at virtually identical temperatures in this system. However, alternative explanations of the experimental findings have also been proposed [22] and we are still far from a final clarification of the microscopic origin of the intriguing dielectric properties of this system. We will discuss in detail the various dielectric results of this system collected by different groups.

In some dielectric investigations of (ET)$_2X$ systems also the typical signatures of relaxor ferroelectricity were found [8,9,10,23]. This phenomenon is usually ascribed to the glasslike freezing-in of short-range clusterlike ferroelectric order. The most prominent system is $\kappa$-(ET)$_2$Cu$_2$(CN)$_3$, having the same molecular arrangement in the ET layers as multiferroic $\kappa$-(ET)$_2$Cu[N(CN)$_2$]Cl. Interestingly, in this material relaxor ferroelectricity, i.e., the absence of long-range polar order, is combined with a spin-liquid state in marked contrast to $\kappa$-(ET)$_2$Cu[N(CN)$_2$]Cl, where both dipolar and spin degrees of freedom order. A very recent example of relaxor ferroelectricity is $\alpha$-(ET)$_2$I$_3$, which will be treated in detail in this Topical Review. In contrast to $\kappa$-(ET)$_2$Cu[N(CN)$_2$]Cl and $\kappa$-(ET)$_2$Cu$_2$(CN)$_3$, CO in this material is well established and recently relaxor behaviour was detected deep in the charge-ordered state [10]. Various dielectric investigations of this system exist and have been interpreted in different ways [10,11,24].

In the first part of this Topical Review, we will briefly introduce typical experimental methods used to investigate the dielectric properties of organic charge-transfer salts (section 2). A large section will be devoted to the rich variety of dielectric phenomena known to occur in these materials and their proper analysis (section 3). Aside of introducing the general features of dielectric relaxation and hopping conductivity, we will provide a short overview of the classes and mechanisms of ferroelectricity. An especially important topic are non-intrinsic dielectric effects as Maxwell-Wagner relaxations, which can lead to the erroneous detection of dipolar relaxation processes and "colossal" values of the dielectric constant ($\varepsilon' \gtrsim 1000$). Section 4 presents experimental results on the two main material classes mentioned above as taken from literature or measured by our own group. A short summary is provided in section 5.

It should be noted that the intention of this work is not to provide an exhaustive review of all dielectric investigations performed on organic charge-transfer salts. Especially, we will not treat SDW or CDW systems, which also have typical dielectric signatures. Instead, we concentrate on systems that are in the focus of recent interest due to their ferroelectric or ferroelectric-like properties, which most likely are of electronic origin.

## 2. Experimental techniques

Most dielectric investigations of charge-transfer salts use "classical" dielectric techniques employing autobalance bridges or frequency-response analyzers to investigate their dielectric response. Especially for insulating materials, the latter devices are better suited as they allow the detection of much lower conductance values, down to an astounding $10^{15}$ $\Omega^{-1}$ (for more details on these techniques, see [25]). Thus these devices by far surpass the lowest conductivities detectable by typical dc experiments and enable following the charge-transport behaviour down to much lower temperatures. These methods typically cover frequencies from the sub-Hz range up so several MHz. This is fully sufficient, e.g., to characterize ferroelectric states and to distinguish between different classes of ferroelectrics as treated in section 3.3 or to investigate relaxational processes. For these dielectric measurements, usually two metallic contacts have to be applied to the samples, as the devices are designed to operate in twopoint or pseudo-fourpoint contact geometry. Sometimes the application of contacts can be challenging, especially if the sample material is available in the form of tiny crystals only. Metallic contacts can be applied, e.g., by evaporation or sputtering of thin films (gold or silver in most cases), requiring the fabrication of suitable evaporation masks. However, generally the application of suspensions of metallic particles as silver paint or graphite paste also leads to satisfactory results. In this case, care should be taken to avoid a dissolving or decomposition of the organic materials by the solvent of the suspension.

For thin platelike samples, often found, e.g., for single crystals of two-dimensional materials, the contacts can be applied at opposite sides of the crystals, thus forming a parallel-plate capacitor. This corresponds to an electric-field direction perpendicular to the planes of the crystalline lattice. For in-plane measurements, coplanar contact geometry can be used, however, making the determination of absolute values of the dielectric quantities difficult. Alternatively, contacts can be applied forming "caps" around opposite ends of the sample. This also works for needle-like crystals as often found for one-dimensional materials. Especially the latter type of samples often suffers from cracks occurring during cooling, which can make the determination of absolute values of the dielectric quantities difficult [1,4]. Sometimes it can be avoided by slow cooling of the samples [1].

Usually, the capacitance and conductance of the samples are measured by the employed devices, from which the dielectric permittivity (real and imaginary part, $\varepsilon'$ and $\varepsilon''$, respectively) and/or the complex conductivity ($\sigma' + i\sigma''$) can be determined by textbook formulae involving the geometry of the sample. Alternatively, also the impedance, i.e., the complex resistance can be determined, from which the resistivity ($\rho' + i\rho''$) can be deduced. Generally, permittivity, conductivity, and resistivity (all being complex quantities) are interrelated by simple formulae (the most common one is $\varepsilon'' = \sigma'/(2\pi\nu\varepsilon_0)$, where $\varepsilon_0$ is the permittivity of vacuum) and in principle provide the same information. However, different dielectrically active processes often are differently emphasized in these quantities, which sometimes makes their separate discussion helpful [26]. For example, it seems natural that plotting the real part of the conductivity is best suited to obtain information on the charge transport in the material, while the dielectric loss is best suited to reveal dipolar relaxation processes.

At frequencies above some MHz, different techniques are employed. For measurements up to several GHz, a coaxial reflectometric method is the most common technique. Here, the sample is mounted at the end of a coaxial line, bridging the inner and outer conductors via a specially designed sample holder [25,27]. Still two metallic contacts on the sample are necessary. An impedance analyzer or network analyzer connected to the other end of the line determines the impedance of the sample either by measuring the complex reflection coefficient of this assembly or by an I-V method based on the direct relation of the voltage-current ratio to the impedance. Proper calibration using at least three standard samples is necessary to eliminate the influence of the coaxial line. The contribution of the sample holder is mainly given by an additive inductance contribution and has to be eliminated by an additional calibration, where the sample is replaced by a short of similar geometry.

Various other techniques are available for this frequency range beyond MHz, also extending into the microwave range up to several tens of GHz [25,28,29,30,31]. For example, resonance methods can be used, where damping and phase shift of a cavity or other resonator with the sample inserted is measured by a network analyzer (no contacts are needed here). Network analyzers can also be employed to measure the transmission of various types of waveguides (e.g., coplanar or striplines) that is damped by the sample material. While in these methods the electromagnetic waves are guided in coaxial of other transmission lines, at even higher frequencies unguided wave methods have to be used. Beyond about 50 GHz up to the THz range, the transmission or reflection of the wave can be detected by quasi-optical techniques [32]. In most cases, both the absolute value and phase shift are measured allowing the direct calculation of the dielectric properties. At even higher frequencies, commercial infrared and optical spectrometers are available. In the present work, we do not treat results obtained by quasi-optical methods or infrared and optical spectroscopy.

## 3. Dielectric phenomena and analysis

Dielectric spectroscopy, even when assuming its "classical" definition restricted to measurements up to about 1 GHz only, is able to detect numerous, very different physical phenomena in condensed matter [33,34]. Here we only treat those known to occur in the organic charge-transfer salts

*3.1 Dielectric Relaxation*

The term "dielectric relaxation" usually denotes reorientational processes in condensed matter that can be detected by dielectric spectroscopy. To enable a coupling to the electrical field, the reorientation must be accompanied by the motion of charges. The classical case is the rotation of dipolar molecules as schematically indicated in the left part of figure 1. The detailed investigation of such processes in a broad frequency range is a standard method for the investigation of the molecular dynamics in molecular solids, supercooled liquids and glass forming systems [33,34]. However, local hopping processes in double- or multiple-well potentials, which formally corresponds to the reorientation of a dipolar moment, also can lead to the typical signatures of dipolar relaxation. Such processes are known, e.g., to sometimes occur in hydrogen-bonded materials, when the proton on an H-bond can hop between two positions and, at low temperatures, either can order ferroelectrically or show glassy freezing [35,36].

The signature of a relaxation process in dielectric spectroscopy is a peak in the loss $\varepsilon''$ and a step in $\varepsilon'$. These features show up in both frequency- and temperature-dependent plots of the complex permittivity as schematically indicated in figures 1(b) - (d). The occurrence of a step in $\varepsilon'(\nu,T)$ is usually made plausible assuming that the dipoles cannot follow the ac electrical field at high frequencies or low temperatures. This leads to a reduction of polarization and, thus, a steplike decrease of $\varepsilon'$ from the static dielectric constant $\varepsilon_s$ to the significantly lower high-frequency limit of $\varepsilon'$. The latter is denoted as $\varepsilon_\infty$ and arises from the ionic and electronic polarizability of the material. The peak in the loss can be interpreted assuming maximum absorption of field energy when the ac-field frequency matches the reorientation frequency of the dipoles. It should be noted, however, that these simple pictures describing the interaction of the ac field with dipolar motion are somewhat oversimplified. Detailed theoretical treatments, mostly founded on Debye's pioneering works on the relaxation dynamics of polar molecules [37], can be found in various textbooks [26,33,38]. Interestingly, the experimentally observed peak and step-widths often are broader than expected within the Debye theory. A common parameterization of the spectra is achieved by the Havriliak-Negami (HN) formula [39]:

$$\varepsilon^* = \varepsilon_\infty + \frac{\varepsilon_s - \varepsilon_\infty}{[1+(i\omega\tau)^{1-\alpha_{HN}}]^{\beta_{HN}}} \qquad (1)$$

Here $\tau$ is the relaxation time characterizing the dynamics of the reorientational motion. The parameter $\alpha_{HN} < 1$ leads to a symmetric and $\beta_{HN} < 1$ to an asymmetric broadening of the loss peaks, compared to the Debye case, where $\alpha_{HN} = \beta_{HN} = 1$. The Cole-Cole (CC) [40] and the Cole-Davidson (CD) [41,42] functions, where $\beta_{HN} = 1$ or $\alpha_{HN} = 0$, respectively, are special cases of this equation. The broadening is usually ascribed to a distribution of relaxation times caused by disorder, a concept that seems especially plausible for glassy or disordered matter [43,44].

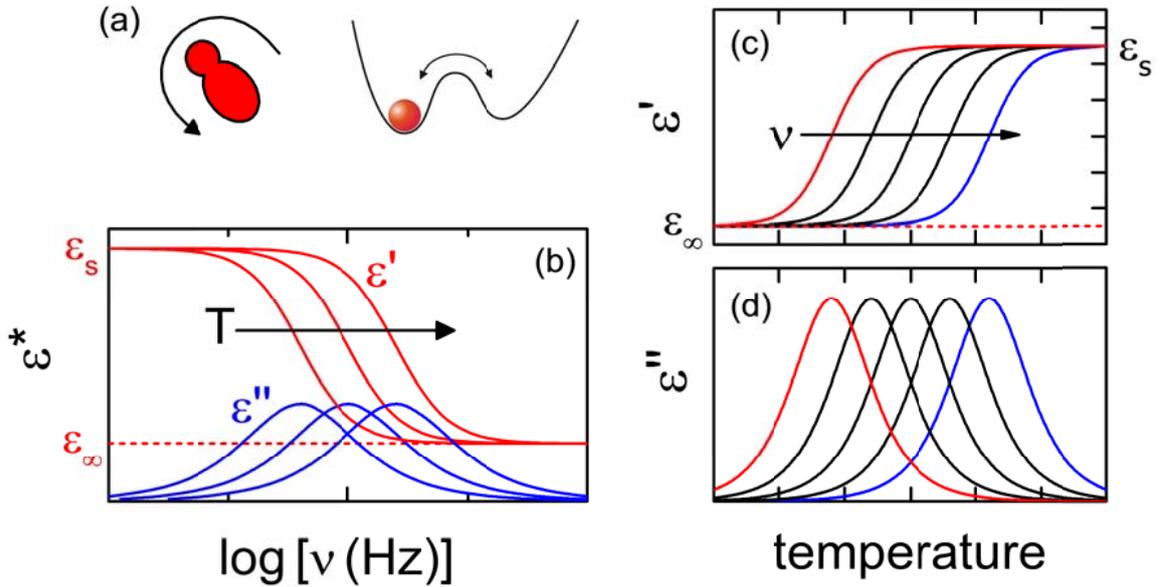

**Figure 1.** (a) Schematic representation of two processes that can lead to dielectric relaxation: molecular reorientation and hopping of charged particles in a double-well potential. (b) Schematic plot of the frequency dependence of the dielectric constant and loss for different temperatures for a dielectric relaxation process. (c) and (d) Temperature dependence of dielectric constant (c) and loss (d) at different frequencies. The dashed lines in (b) and (c) indicate the high-frequency limit of the dielectric constant.

The temperature-induced shift of the step observed in the $\varepsilon'(\nu)$ spectra or of the peak in $\varepsilon''(\nu)$ [figure 1(b)], mirrors the slowing down of the dipolar motion at low temperatures. From the frequency $\nu_p$ of the peak in $\varepsilon'(\nu)$ or of the point of inflection in $\varepsilon''(\nu)$, the relaxation time $\tau$ can be determined via $\tau \approx 1/(2\pi\nu_p)$ (for the Debye and CC cases, this relation is exact [38]). In the simplest case, $\tau(T)$ shows thermally activated behaviour, $\tau = \tau_0 \exp[E/(k_B T)]$, where $E$ denotes the energy barrier for reorientational motion and $\tau_0$ is an inverse attempt frequency. Nevertheless, often deviations from this formula are found. A common empirical parameterization is the Vogel-Fulcher-Tammann (VFT) function [45,46,47]:

$$\tau = \tau_0 \exp\left[\frac{B}{T-T_{VF}}\right] \qquad (2)$$

Here $B$ is an empirical constant and $T_{\text{VFT}}$ corresponds to the divergence temperature of $\tau_{\text{peak}}$. This function is known to provide a good parameterization of the slowing down of molecular motion of disordered matter [33,34]. However, it often is also used to describe the glass-like freezing of dipolar order in the so-called relaxor ferroelectrics (see, e.g., [48,49,50]), treated in section 3.3, and thus also is relevant for various organic charge-transfer salts (e.g., [8,9]). Aside of relaxor ferroelectrics, some canonical ferroelectric materials also are known to exhibit dielectric relaxation behaviour. This is especially the case for the so-called order-disorder ferroelectrics, discussed in section 3.3, and also found in some organic salts (e.g., [3,4]). There the temperature dependence of the relaxation time shows critical behaviour.

Finally, we want to mention that relaxational dielectric response can also be observed in CDW systems, which also may be relevant for some organic charge-transfer salts (e.g., [51,52,53]). Their dielectric behaviour shows two characteristic features: A resonance mode at GHz frequencies arising from the CDW that is pinned at defects and a relaxation mode at kHz-MHz with giant values of the static dielectric constant [54]. As suggested by Littlewood [55], the low-frequency relaxation mode in this class of materials may arise from the screening of the pinned CDW by the normal electrons that do not participate in the CDW.

*3.2 Hopping charge transport*

In electronic conductors, disorder can lead to a localization of the charge carriers (electrons or holes). The disorder may be caused by an amorphous structure (e.g., amorphous Si) or doping (substitutional disorder). Even in nominally pure crystals, slight deviations from stoichiometry or lattice imperfections can occur. Charge transport of localized charge carriers takes place via hopping from one minimum in the disordered energy landscape to another as schematically indicated in the inset of figure 2. The hopping processes can be thermally activated or may occur via tunneling. Aside of classical examples as amorphous or heavily doped semiconductors, hopping conductivity was also found in a large variety of electronically correlated materials (e.g., [56,57,58,59,60,61,62]), also including several organic charge-transfer salts (e.g., [7,63,64,65]). There are various theoretical treatments of hopping charge transport (see, e.g., [66,67,68,69]), the most prominent one being Mott's variable-range hopping (VRH) model, assuming phonon-assisted tunneling processes [70]. Distinct predictions are made for the temperature dependence of the dc and ac conductivity. Best known is the VRH formula for the dc conductivity:

$$\sigma_{\text{dc}} = \sigma_0 \exp[-(T_0/T)^{1/\gamma}] \qquad (3)$$

For isotropic charge transport, $\gamma = 4$ is predicted. Values of $\gamma = 3$ and 2 arise for VRH conduction in two and one dimensions, respectively [70,71], which may be relevant for the low-dimensional charge transport in organic charge transfer salts. A value of $\gamma = 2$, however, can also arise in three dimensions when additional Coulomb interaction between the charge carriers is taken into account [72].

The characteristic signature of hopping conductivity in the frequency dependence of the complex conductivity is a power law with an exponent $s < 1$, usually termed "universal dielectric response" (UDR) as schematically shown in figure 2 [26]:

$$\sigma' = \sigma_{\text{dc}} + \sigma_0 \nu^s \qquad (4a)$$
$$\sigma'' = \tan(s\pi/2)\,\sigma_0\,\nu^s + \varepsilon_\infty\,\varepsilon_0\,2\pi\nu \qquad (4b)$$

The last term in equation (4b) accounts for the contribution of $\varepsilon_\infty$, using the relation $\sigma'' = 2\pi\nu\varepsilon_0\varepsilon'$ (it causes the increasing slope observed in $\sigma''$ at the highest frequencies in figure 2). $\sigma_0$ is a prefactor and $\varepsilon_0$ is the permittivity of free space. The temperature dependence of the experimentally determined $s$ can be used to check the applicability of a specific hopping model for the investigated material [66,67,68]; e.g., for VRH, a temperature independent $s \approx 0.8$ is expected [67,70]. In accord with experimental observations, most models agree in predicting such a power law for the conductivity. However, it should be noted that often slight deviations from a pure $\nu^s$ law are expected, corresponding to a weak frequency dependence of the exponent *s* [66,67]. Generally, the ac part of $\sigma'(\nu)$ in equation (4a) exhibits a weaker temperature dependence than the dc part [66,67], which often is thermally activated or follows equation (3).

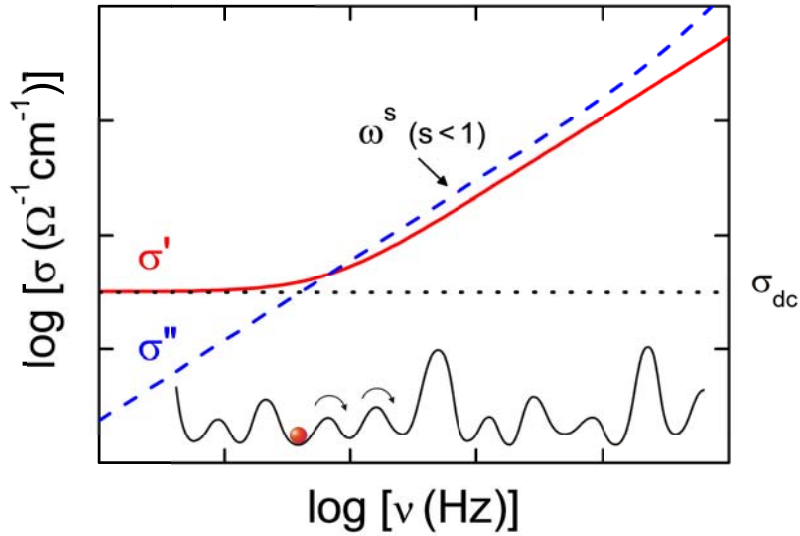

**Figure 2.** Schematic representation of the UDR usually observed in spectra of $\sigma'$ (solid line) and $\sigma''$ (dashed line) for hopping charge transport [equations (4a) and (4b)]. The dotted line indicates the dc conductivity, leading to a low-frequency plateau in $\sigma'(\nu)$. The inset schematically indicates the hopping of a localized charge carrier in a disordered energy landscape.

The UDR also leads to contributions in the permittivity via the relations $\varepsilon' = \sigma'' / (2\pi\nu\varepsilon_0)$ and $\varepsilon'' = \sigma' / (2\pi\nu\varepsilon_0)$. The dc conductivity produces a $1/\nu$ law in $\varepsilon''(\nu)$ and the ac part of equation (4) corresponds to a $\nu^{s-1}$ power law in both, $\varepsilon'(\nu)$ and $\varepsilon''(\nu)$. Thus, these conductivity contributions lead to low-frequency divergences in both parts of the permittivity that can obscure other contributions to the spectra, e.g., from relaxation processes.

*3.3 Ferroelectricity*

The most prominent methods to identify ferroelectricity in organic charge transfer salts certainly are dielectric spectroscopy, which allows detecting anomalies in the temperature dependence of the dielectric constants, pyrocurrent experiments to monitor the onset of macroscopic ferroelectric polarization and ferroelectric hysteresis experiments, i.e. measurements of the polarization *P* versus an external electric field *E*, to observe its switchability. In many cases, and this specifically is true for organic charge transfer salts, the identification of a polar ground state is not simple and hampered by conductivity contributions and/or extrinsic Maxwell-Wagner like relaxation phenomena (for details see section 3.4). Hence, to begin with it seems useful to define ferroelectricity and to identify prototypical experimental signatures. There exists a number of time-honoured review articles and books on ferroelectricity, including, Devonshire [73], Cochran [74], Jona and Shirane [75], R. Blinc and B. Zeks [76] and Lines and Glass [77] to name only a few of them.

As starting point of our discussion, we provide the standard characterization of ferroelectricity: A ferroelectric is by definition a crystalline solid with two properties: It belongs to a polar crystal point group and the polarity can be reversed by an external electric field. All crystals belonging to ten of the 32 point groups are polar, but only in a small fraction of these the polarity can be reversed. Switchability means that the relationship between the dielectric displacement *D* and the electric field *E* is hysteretic and in canonical ferroelectrics hysteresis loops *P*(*E*) are obtained in alternating electric fields. However, in many cases at low temperatures domains are large and domain walls can hardly be moved by realistic external electric fields (less than breakdown field), while at high temperatures charge carriers become mobile resulting in hysteresis curves dominated by Ohmic losses, again hampering switchability. Hence, in real systems macroscopic and intrinsic polarization sometimes is hard to detect, which is specifically true for many organic polar crystals.

At a critical temperature $T_c$, a ferroelectric material undergoes a structural phase transition from a high-temperature, high-symmetry and paraelectric phase into a low-symmetry ferroelectric phase. The low-symmetry phase is characterized by the occurrence of spontaneous polarization $P$ which is reversible under the influence of external electric fields. Phase transitions in ferroelectrics are usually connected with mostly minor rearrangements of a few atoms only: In ferroelectric perovskites of stoichiometry $ABO_3$, the $B$ ions, octahedrally coordinated by six $O^{2-}$ ions, move off-center resulting in a long-range ordered polar ground state. In hydrogen-bonded ferroelectrics, at high temperatures and in the paraelectric phase the protons are statistically distributed in a double-well potential located between two neighbouring oxygen ions, but undergo long-range order with the preference of one site in the ferroelectric state. Molecular compounds with permanent dipole moments, at high temperatures often exhibit orientationally disordered phases, but reveal long range order of the dipole moments in the low-temperature polar phase. The fact that one class of crystals exhibits a high-symmetry structure with no permanent dipole moments at high temperatures and undergoes polar order with permanent dipole moments via local shifts of ions and a second class of crystals is characterized by the existence of permanent dipole moments, which in the high temperature phase are statistically disordered and undergo polar order by aligning these dipolar moments, is frequently used as a classification scheme of ferroelectrics into classes of displacive *vs*. order-disorder phase transitions, respectively: In the high-temperature paraelectric phase, displacive ferroelectrics have no permanent dipole moments, while order-disorder ferroelectrics have permanent dipoles which are disordered with respect to site and time.

### 3.3.1 Displacive ferroelectrics

In displacive ferroelectrics one specific vibrational mode, a dipolar-active transverse optical phonon, at the zone centre of the Brillouin zone becomes soft and finally condenses at a critical temperature. This freezing-in of the long-wavelength ionic displacements results in an off-symmetry position of the cation in the cage of surrounding anions establishing long-range ferroelectric polarization. For the case of a ferroelectric with a single soft mode, the relation of the dielectric constants to the optical eigenfrequencies are given by the Lyddane-Sachs-Teller (LST) relation $\varepsilon(0)/\varepsilon_\infty = (\omega_{LO}/\omega_{TO})^2$. Here $\varepsilon(0)$ is the static and $\varepsilon_\infty$ the high-frequency dielectric constant; $\omega_{LO}$ and $\omega_{TO}$ are the longitudinal and transversal zone-centre optical phonons, respectively. The LST relation describes the so-called polarization catastrophe, where the softening of the mode is connected with the divergence of the static dielectric constant. The validity of the LST relation has been documented by Cowley via inelastic neutron scattering in $SrTiO_3$ [78].

The divergence of the static dielectric constant is an important signature of displacive ferroelectrics and has been observed in many systems, specifically in many perovskites with polar ground states. Figure 3(a) shows an illuminating example, where the static dielectric constant of $BaTiO_3$ diverges at $T_c$ [79]. At $T_c$ the high-symmetry cubic structure transforms into a low symmetry tetragonal polar state with the occurrence of spontaneous ferroelectric polarization. The temperature dependence of the inverse dielectric constant in the paraelectric phase shows that the temperature dependence of the real part of the dielectric constant can be perfectly well described by a Curie-Weiss law

$$\varepsilon' = C / (T - T_{CW}) \qquad (5)$$

with $C$ the Curie constant determined by the number of dipoles per volume and the size of the dipole moment and $T_{CW}$ the Curie-Weiss temperature which gives an average interaction strength between neighbouring dipoles. In many cases, and this is also documented in figure 3(a), the Curie-Weiss temperature is of order of the ferroelectric ordering temperature.

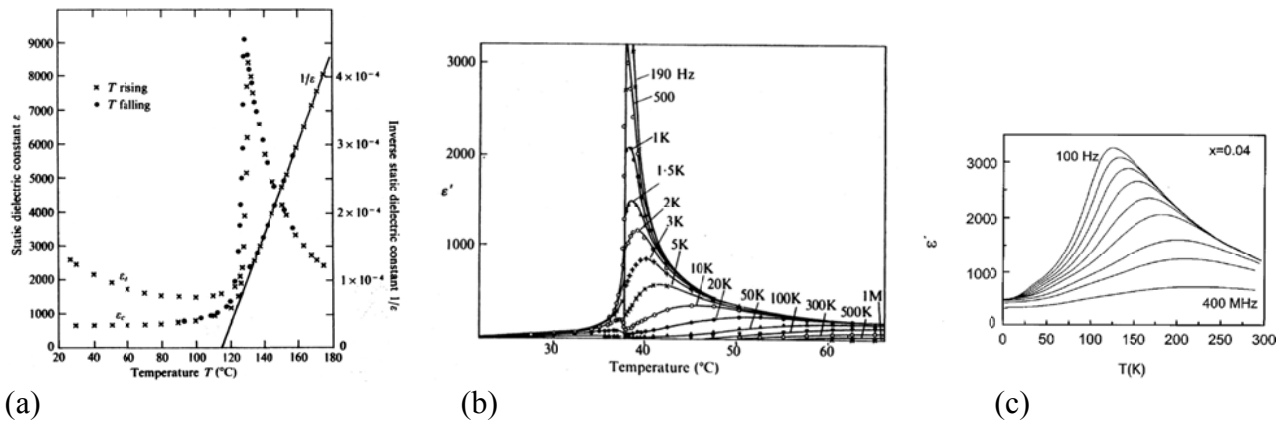

(a) (b) (c)

**Figure 3.** Three typical examples taken from literature [79,81,86] for the temperature dependence of the dielectric constant of materials belonging to the three classes of ferroelectrics treated in sections 3.3.1 - 3.3.3. (a) BaTiO$_3$ (reprinted with permission from [79]. Copyright 1965 by AIP Publishing), (b) AgNa(NO$_2$)$_2$ (reprinted from [81] with kind permision from the Physical Society of Japan) and (c) (Sr$_{1-1.5x}$Bi$_x$)TiO$_3$ ($x$ = 0.04) (reprinted from [86]. Copyright 1999 by the American Physical Society). For more details on these figures, see the original publications.

*3.3.2 Order-disorder ferroelectrics*

As outlined above, order-disorder ferroelectrics are characterized by permanent but disordered dipole moments already in the paraelectric phase. The temperature dependence of their dielectric response is often dominated by strong dispersion effects at audio and radio frequencies. This dispersion effects result from the slowing down of molecular reorientations within a double- or multi-well potential on decreasing temperatures. In most ferroelectrics where dispersion effects appear at low frequencies, the dielectric response is almost monodispersive and can be approximately described by a Debye relaxation with one well-defined mean relaxation time. Illuminating examples of ferroelectric order-disorder transitions have been observed, e.g., in NaNO$_2$ [80] or in AgNa(NO$_2$)$_2$ [81], in which the NO$_2^-$ ionic groups carry a permanent electric dipole moment. At high temperatures, in the paraelectric phase the molecular units undergo fast reorientations between equivalent crystallographic directions, while they reveal long-range order below the ferroelectric transition temperature. A representative dielectric measurement is shown in figure 4(b) [81]. Here the real part of the dielectric constant of AgNa(NO$_2$)$_2$ is plotted as function of temperature for measuring frequencies between 190 Hz and 1 MHz. It is important to note that, while at low frequencies a well-defined and sharp anomaly is visible, almost no dielectric anomaly appears at high frequencies. It is also important to recognize that at the critical temperature the relaxational processes are abruptly impeded and that dispersion effects, i.e. the shift of the maxima of the dielectric constant for different frequencies, appear in a narrow temperature range. The shift of the maximum between 190 Hz and 10 kHz is less than 10 K compared to an ordering temperature of 390 K. It has been found, that in many cases the temperature dependence of the mean relaxation time cannot be approximated by a simple thermally activated Arrhenius behaviour [10].

*3.3.3 Relaxor ferroelectrics*

Relaxor ferroelectrics can best be described as ferroelectrics with a smeared-out diffusive phase transition extended over a finite temperature range with high values of the dielectric constant. They occur in disordered solids, in particular in solid solutions. Probably the best known relaxor system is the pseudo-cubic perovskite PbMg$_{1/3}$Nb$_{2/3}$O$_3$ (PMN) [82]. All relaxors display large dispersion effects close to the maximum permittivity, which are reminiscent of those found in orientational glasses [83]. Specifically three features in the dielectric response are different to what is observed in canonical ferroelectrics: The transition as evidenced by the temperature dependence of the dielectric constant is clearly rounded with a broad maximum at $T_m$. The smeared-out peak is a function of frequency with characteristic dispersion effects appearing below $T_m$. In most cases the relaxation can also not be described by a simple monodispersive Debye relaxation but is indicative of a distribution of relaxation times. This polydispersivity of the dipolar relaxation is a fingerprint of disorder and cooperativity. The $P(E)$ hysteresis loops can be characterized as so-called slim loops, probably evidencing the freezing and reorientation of nano-scale clusters contrasted to the occurrence of macroscopic domains in conventional ferroelectrics. Reviews on relaxor ferroelectrics have been published by Cross [84] and Samara [85]

The dielectric behaviour revealed by figure 3(c) [86] is typical for relaxor ferroelectrics. The observed frequency-induced shift of the peak temperatures in the dielectric constant can usually be described by the VFT law, equation (2), again typical for relaxor ferroelectrics (see, e.g., [48,49,50]). In this type of analysis, the temperature in equation (2) corresponds to the peak temperature in $\varepsilon'(T)$ and the relaxation time is calculated from $\tau_{peak} = 1/(2\pi\nu)$, where $\nu$ is the frequency of the applied ac field. It should be noted that, according to the Debye theory [37,38], the mean relaxation time $\tau$ should be related to the dielectric-loss peak and that no peak occurs in $\varepsilon'(T)$ for conventional dipolar relaxation processes [cf. figure 1(c)]. Nevertheless, evaluating the peaks in the real part and comparing $\tau(T)$ derived from it with the VFT equation is common practice for relaxor ferroelectrics. It also seems important to note that in relaxor ferroelectrics in the paraelectric phase for temperatures above the susceptibility maximum, the dipolar susceptibility often deviates from Curie-Weiss behaviour, eq. (5), characteristic for conventional ferroelectrics. It was argued that these deviations arise due to short-range correlations between nano-scale polar regions and that these correlations at high temperatures are the precursors of freezing of the polarization fluctuations into a glassy low-temperature state [87].

*3.3.4 Spin-driven and electronic ferroelectricity*

In the last decade a number of magnetic compounds have been identified, in which ferroelectricity is induced at the onset of magnetic order. The most prominent examples of these spin-driven ferroelectrics are multiferroic perovskite manganites [88]. In these compounds, just at the onset of helical spin order weak and improper ferroelectricity is induced, an effect that has be explained by an inverse Dzyaloshinskii-Moriya mechanism [89].

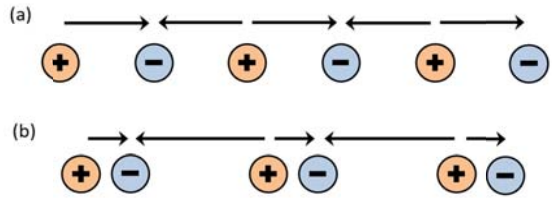

**Figure 4.** Schematic representation of charge ordering without bond order, not leading to polar order (a), and of CO accompanied by bond order (dimerization), generating a net polarization (b) [21]. The arrows indicate the individual dipolar moments, which do not compensate for case (b).

In a variety of materials also charge ordering can induce ferroelectricity [21]. An illuminating example is given in figure 4. Figure 4(a) shows a one-dimensional chain with complete CO, i.e. with opposite charges at equidistant positions. Within this chain the dipole moments, indicated by arrows, exactly compensate with no residual polarisation. If, however, this CO is accompanied by bond order, a net polarization appears in the bond- and site-ordered phase [figure 4(b)]. The observation of CO is a common phenomenon in many transition-metal compounds, with the Verwey transition in magnetite being the most prominent example [90], but is also often observed in low-dimensional organic compounds as outlined below. The coexistence of site and bond order can result purely from crystal symmetry, can be induced by dimerization processes of neighbouring ions or molecules, but also can result from magnetic exchange interactions.

*3.4 Non-intrinsic effects*

The most prominent non-intrinsic contribution to the dielectric response of materials is the long-known Maxwell-Wagner polarization [91,92] sometimes also termed space-charge effect. It arises from charge accumulation at interfaces and can give rise to dielectric constants of extreme magnitude (sometimes termed "colossal dielectric constants" [93,94,95,96]). Any kind of interfaces in the sample can lead to the apparent detection of very high values of the dielectric constant because usually they behave as very thin parallel-plate capacitors, thus leading to high capacitances. For example, such effects can occur at the surface of the sample, e.g., at the electrode-sample interface, giving rise to so-called electrode polarization. This is especially relevant for semiconducting samples and, thus, also for many organic charge-transfer salts. When applying metallic contacts at a semiconducting sample, depletion layers of low carrier concentration may arise at the sample surfaces due to the formation of Schottky diodes at the metal-sample interfaces. Surface layers with ill-defined stoichiometry at the surface of the sample are other possible reasons for surface-related Maxwell-Wagner effects. Moreover, internal interfaces can arise too. Common examples are the grain boundaries of ceramic samples or planar crystal defects like twin boundaries [93].

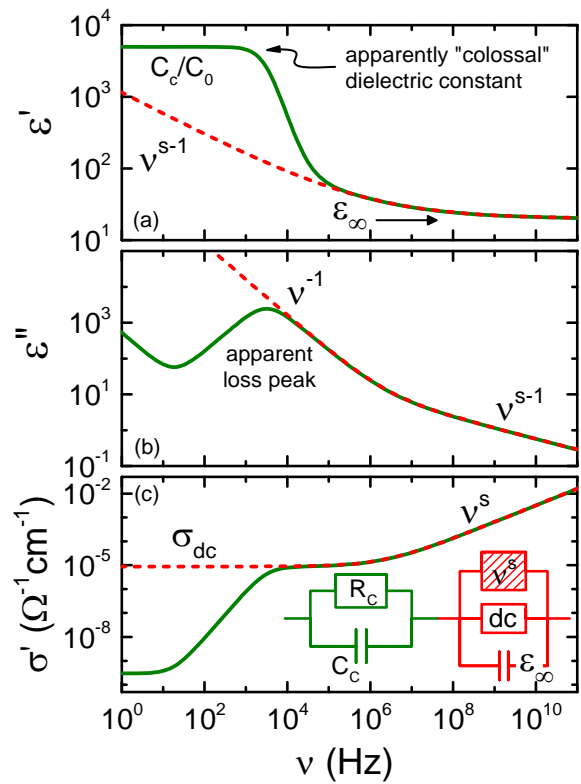

**Figure 5.** Schematic spectra of the real and imaginary part of the permittivity [(a) and (b), respectively] and of the conductivity (c) as resulting from the equivalent circuit shown in (c) [94]. Solid lines: overall response. Dashed lines: intrinsic bulk behaviour including dc conductivity $\sigma_{dc}$, a frequency-independent dielectric constant $\varepsilon_\infty$ and a UDR contribution due to hopping conductivity, as given by equations (4a) and (4b).

Maxwell-Wagner effects generally lead to a strong frequency-dependence of the dielectric properties, even if the investigated material exhibits no intrinsically frequency-dependent microscopic processes. This can be well understood by modelling the different regions of the sample (usually one or several thin interface regions and the bulk) by an equivalent circuit. The simplest case is a parallel RC circuit for the interface layer, connected in series to the bulk element, which, e.g., can be an RC circuit with intrinsic frequency dependence as indicated in the inset of figure 5(c). The interface layers usually have lower conductivity and higher capacitance than the bulk. Therefore, at high frequencies the capacitor caused by the interface becomes shorted and the intrinsic bulk behaviour is observed. Such behaviour is schematically indicated in figure 5, where the dashed lines indicate the intrinsic bulk response and the solid lines the measured spectra [94]. At low frequencies, a Maxwell-Wagner relaxation process arises that exhibits the same spectral signatures as an intrinsic relaxation process (section 3.1). For $\nu \to 0$ the dielectric constant assumes apparently colossal values because here the high capacitance of the thin interface layer dominates the measured $\varepsilon'$. This happens despite the true dielectric constants of both the interface and the bulk are not colossal at all (for a more detailed treatment, see [96]). The curves in figure 5 were calculated assuming UDR behaviour caused by hopping conductivity for the intrinsic dielectric response of the sample material. For an intrinsic relaxation process (section 3.1), a second (intrinsic) set of relaxation features would show up in the spectra at high frequencies (at least if the Maxwell-Wagner and intrinsic relaxation are well separated in frequency). However, sometimes several successive Maxwell-Wagner relaxations also can occur in the spectra. This can happen, e.g., in ceramic samples, where grain boundaries and contact-related Schottky diodes represent two different types of interfaces [96,97,98].

Distinguishing MW relaxations from intrinsic bulk ones is an essential task. A check for contact effects can be performed by a variation of the contact material (e.g., silver paint, carbon paste, sputtered or evaporated contacts), which should affect the contact-dominated regions of dielectric spectra [94,95,96]. Another possibility is the variation of the sample geometry, especially of the area-to-thickness ratio. The enhanced low-frequency dielectric constant for a Maxwell-Wagner relaxation arises from the fact that $\varepsilon'$ is calculated via $C'/C_0$. Here $C'$ is the capacitance measured by the device (which is large for a thin interface layer) and $C_0 = A/d\, \varepsilon_0$ is the empty capacitance of the sample, which depends on the surface-to-thickness ratio $A/d$. Electrode effects are surface effects. Thus, if $C'$ is completely dominated by electrode polarization [usually the case for low frequencies; cf. figure 5(a)], the measured $C'$ will only be determined by the surface and not by the thickness of the sample. Therefore, when measuring two samples with different $A/d$, their calculated dielectric constants would markedly differ, which would not be the case for an intrinsic bulk effect. The best method to exclude grain boundary contributions in the dielectric response is measuring good single-crystalline samples. Alternatively, ceramic samples subjected to different sintering procedures or prepared from differently ground powders, leading to differently sized grains, should be investigated and the results compared to unequivocally reveal grain-boundary effects.

It should be noted that, of course, all the effects treated in chapter 3 also can simultaneously occur in a sample. This is especially likely in the organic charge-transfer salts as even those classified as "insulating" have non-negligible conductivity. The latter contributes to the dielectric spectra in a direct way, e.g., by the divergence of the permittivity mentioned in section 3.2 and also indirectly as samples that are not completely insulating are prone to the formation of electrode polarization as discussed in the previous paragraphs. It often is difficult to deconvolute the different contributions in the dielectric spectra. Sometimes impedance plots or the subtraction of conductivity contributions are employed to analyse such data. Nowadays, assuming proper equivalent circuits and making simultaneous least-square fits of both $\varepsilon'(\nu)$ and $\varepsilon''(\nu)$ or of $\varepsilon'(\nu)$ and $\sigma'(\nu)$ is state of the art for a proper analysis of such data. However, even then it may prove difficult to unequivocally determine the parameters of the different processes in the material, especially if their contributions in the spectra are strongly superimposed by each other or are obscured by non-intrinsic Maxwell-Wagner effects.

## 4. Experimental results and interpretation

*4.1 One-dimensional ferroelectric charge-transfer salts*

*4.1.1 (TMTTF)$_2$X*

Concerning their dielectric properties, among the one-dimensional organic charge-transfer salts especially those from the (TMTTF)$_2$X family are of interest (typical anions $X$ are Br, PF$_6$, AsF$_6$, SbF$_6$, BF$_4$, ReO$_4$, and SCN [1]). They feature nearly planar organic TMTTF (tetramethyltetrathiofulvalene) molecules that are stacked along one direction, forming chainlike structures with rather strong dimerization. More details on these systems can be found, e.g., in several review articles [1,12,99]. Their dielectric properties were investigated in great detail in a number of works by Nad, Monceau, and coworkers (see, e.g., [1,2,3,100,101,102,103,104]. CO arising from charge disproportionation between neighbouring TMTTF molecules is a common phenomenon in these materials [1,12] as was revealed, e.g., by NMR [105,106]. As expected, the CO transition also clearly affects the dc conductivity of these systems and most of them exhibit an "abrupt drop of conductance" [1] below $T_{CO}$. As an example, figure 6 shows the conductivity of $(TMTTF)_2AsF_6$ at 1 kHz as reported in ref. [102], representing a good approximation of the dc behaviour. At $T_{CO}$, the dielectric constant $\varepsilon'(T)$ of this system exhibits a sharp maximum, indicating ferroelectric ordering (figure 7 [1]). Some of these materials also are known to undergo an additional so-called anion-ordering transition at lower temperatures, which has a less dramatic, kink-like signature in the dielectric constant [1].

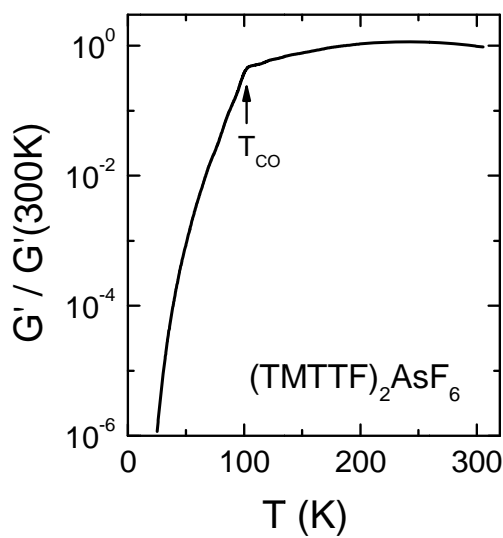

**Figure 6.** Temperature dependence of the scaled conductance of $(TMTTF)_2AsF_6$ at 1 kHz (the data were taken from [102], where they were shown in Arrhenius representation). The arrow indicates the temperature of the CO transition.

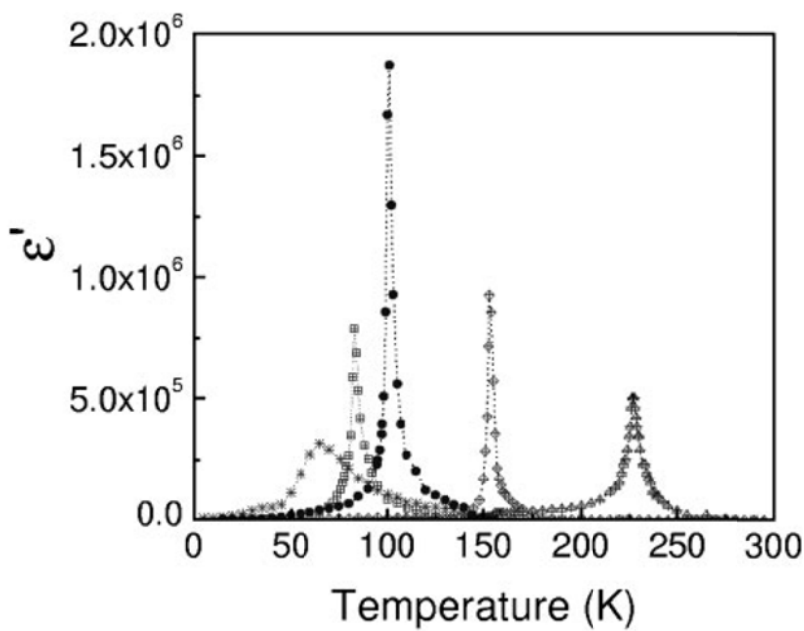

**Figure 7.** Temperature dependence of the dielectric constant at 100 kHz for various members of the $(TMTTF)_2X$ family of one-dimensional charge transfer salts [1]. $X = PF_6$ (stars), $BF_4$ (squares), $AsF_6$ (circles), $SbF_6$ (diamonds), $ReO_4$ (triangles). (Reprinted from [1] with kind permission from the Physical Society of Japan.)

As no significant structural variation accompanies the CO and ferroelectric transition, these systems are prototypical examples of electronic ferroelectrics (cf. section 3.3.4). However, it should be noted that, generally, considering purely electronic effects is likely to be an oversimplification and the counterions may also play some role in forming the net polar moment. An interesting example in this regard is the CO in δ-[EDT-TTF-CONMe$_2$]$_2$X, where the CO is stabilized by the anion shift [107]. For the (TMTTF)$_2$X family, the role of the counterions was considered already in ref. [2] and is also discussed in detail in refs. [108,109,110]. Finally we want to note that within this class of materials, (TMTTF)$_2$SbF$_6$ is an especially interesting case. This system was reported to become ferroelectric below about 154 K [1,2] and to show antiferromagnetic order below 8 K [109,111]. Thus, as recently explicitly pointed out in ref. [112], (TMTTF)$_2$SbF$_6$ is multiferroic. As discussed in refs. [112,113] there seems to be a close interplay of polar and magnetic order in this system, which was also considered to be relevant for multiferroic κ-(ET)$_2$Cu[N(CN)$_2$]Cl [112].

In the following, as a typical example for the (TMTTF)$_2$X systems, we will discuss in some more detail the dielectric reponse of (TMTTF)$_2$AsF$_6$. Figure 8 shows the temperature-dependent dielectric constant of (TMTTF)$_2$AsF$_6$ as measured at different frequencies [4]. At low frequencies, a sharp peak in $\varepsilon'(T)$ is observed, indicating the ferroelectric transition, which occurs at $T_{CO}$ of this system (cf. figure 6). Its amplitude diminishes when the frequency increases. While at low frequencies the peak position is nearly constant, at higher frequencies a slight shift to higher temperatures is observed. In this respect, the dielectric response closely resembles that of typical order-disorder ferroelectrics (section 3.3.2) [77]. Qualitatively very similar behaviour of (TMTTF)$_2$AsF$_6$ as in figure 8 was also reported in ref. [102]. The significantly higher peak values of the dielectric constant found in this work (see also figure 7), points to a strong sample dependence of the dielectric properties. In the inset of figure 8, $1/\varepsilon'(T)$ at the lowest frequency of 30 kHz is shown. It represents a good estimate of the static dielectric constant [4]. The solid line demonstrates Curie-Weiss behaviour [eq. (5)] at temperatures above the ferroelectric transition. Below the transition, for canonical second-order phase transitions another Curie-Weiss law with a slope that is twice as steep is expected as indicated by the dashed line [77]. This is not well fulfilled in this case which may be ascribed to disorder [1,4].

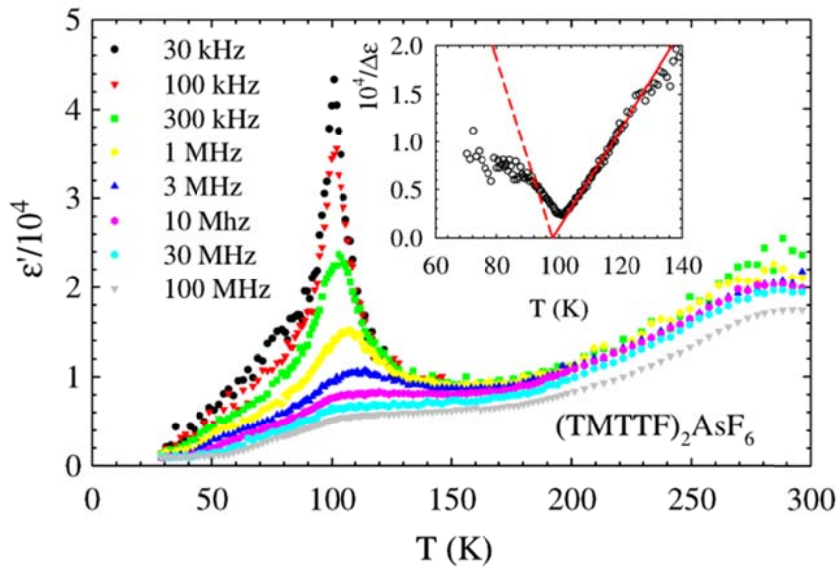

**Figure 8.** Temperature dependence of the dielectric constant of (TMTTF)$_2$AsF$_6$ measured at various frequencies [4]. The inset shows the inverse dielectric constant for the lowest investigated frequency. The solid line indicates a fit of the data at temperatures above the minimum by a Curie-Weiss law. The dashed line demonstrates the behaviour expected for a canonical second-order phase transition with a slope that is twice as steep. (Reprinted from [4], copyright (2006), with permission from Elsevier.)

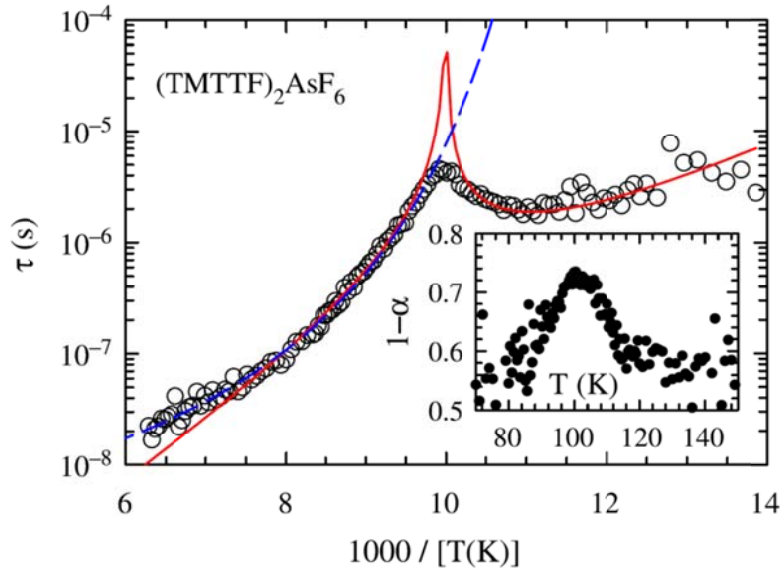

**Figure 9.** Arrhenius plot of the temperature dependence of the dielectric relaxation time in $(TMTTF)_2AsF_6$ [4]. The lines are fits with two alternative models (see [4] for details). The inset shows the temperature dependence of the parameter characterizing the width of the relaxational process observed in the permittivity spectra [$1-\alpha_{HN}$ in eq. (1)] [4]. (Reprinted from [4], copyright (2006), with permission from Elsevier.)

Plots of $\varepsilon'(\nu)$ [3,4] clearly reveal that the frequency dependence observed in figure 8 is of relaxational character. From an evaluation of the spectra of the real and imaginary parts of the permittivity, the corresponding relaxation time can be determined. Its temperature dependence is shown in figure 9 using an Arrhenius representation [4]. A peak in $\tau(T)$ is observed at the ferroelectric transition, in agreement with the findings in [3]. It is known that canonical order-disorder ferroelectrics exhibit such a maximum [77]. In ref. [4] it was considered to arise from critical soft-mode behaviour and in ref. [3] it also was associated with the "softening of the oscillating mode responsible for the observed ferroelectric transition". The inset of figure 9 shows the width parameter $1-\alpha$ of the Cole-Cole function [equation (1) with $\beta_{HN} = 1$] used to fit the permittivity spectra [4]. Obviously, the relaxation-time distribution in $(TMTTF)_2AsF_6$ also reflects the ferroelectric transition.

Finally, we want to mention that very recently, from an analysis of the dielectric modulus, the inverse of the permittivity, a second dynamical process was found in this material [104]. It was detected below $T_{CO}$ and ascribed to "slow oscillations of pinned ferroelectric domains". The modulus representation is quite common in the analysis of dielectric spectra of ionic conductors [114]. Its use for organic charge-transfer salts seems an interesting approach and may lead to new insights on their relaxational behaviour in the future.

*4.1.2 TTF-BA and TTF-CA*

The next system that we want to briefly discuss is tetrathiafulvalene-p-bromanil (TTF-BA). The crystalline structure of this material includes mixed stacks consisting of the $TTF^+$ cations and $BA^-$ anions [5]. A spin-Peierls transition occurs at $T_c \approx 53$ K leading to dimerized $TTF^+BA^-$ stacks. Pioneering dielectric measurements of polycrystalline samples [115] revealed a peak in $\varepsilon'(T)$ at $T_c$ with amplitude of about 15. Moreover, the typical features of relaxational behaviour (section 3.1) were found at higher temperatures (100 - 300 K) with a high static dielectric constant of the order of 1000. The latter was ascribed to thermally activated motions of ferroelectric domain-walls. In a more recent work [5], the occurrence of ferroelectricity in this materials was corroborated by dielectric measurements of single crystals, revealing a sharp peak in $\varepsilon'(T)$ with an amplitude of 200, and by polarization measurements. It should be noted that, in contrast to the (TMTTF)$_2$X systems, the spin-Peierls induced ferroelectricity of TTF-BA arises from a displacement of ions and is not of electronic nature.

An even more interesting system is tetrathiafulvalene-p-chloranil (TTF-CA). As for TTF-BA, its structure also includes mixed stacks of the two molecular species forming it [6,116]. Similar to TTF-BA, the material undergoes a transition into a ferroelectric state, occurring simultaneously with a dimerization of the stacked molecules. As shown in figure 10(b) [6], this transition taking place at $T_c \approx 81$ K leads to a pronounced peak in the dielectric constant. Its right flank follows Curie-Weiss behaviour (inset). However, the microscopic origin of this ferroelectric transition is markedly different compared to TTF-BA. As revealed by figure 10(a) [6], in contrast to TTF-BA the ionicity $\rho$ of TTF-CA markedly increases at $T_c$ from about 0.3 to 0.6. Thus in literature it is also referred to as "neutral-to-ionic" phase transition [117]. Astonishingly, the ionic displacements occurring at $T_c$ cannot explain the experimentally observed polarization. Instead, as theoretically predicted [118], in this material electronic ferroelectricity dominates and in fact the much stronger electronic contribution to the polarization is even antiparallel to the ionic one [6,116]. In early dielectric measurements of this material [119], in addition to the typical peak marking the ferroelectric transition, a huge relaxational mode was observed at $T > T_c$. In that work, the dielectric results were explained by the dynamics of domains and domain walls. In figure 10(b) a less prominent but still significant relaxation-like response is observed at the right flank of the ferroelectric peak. It was ascribed to the dynamics of "locally pinned polar domains". For a detailed comparison of the properties of TTF-BA and TTF-CA, see ref. [116].

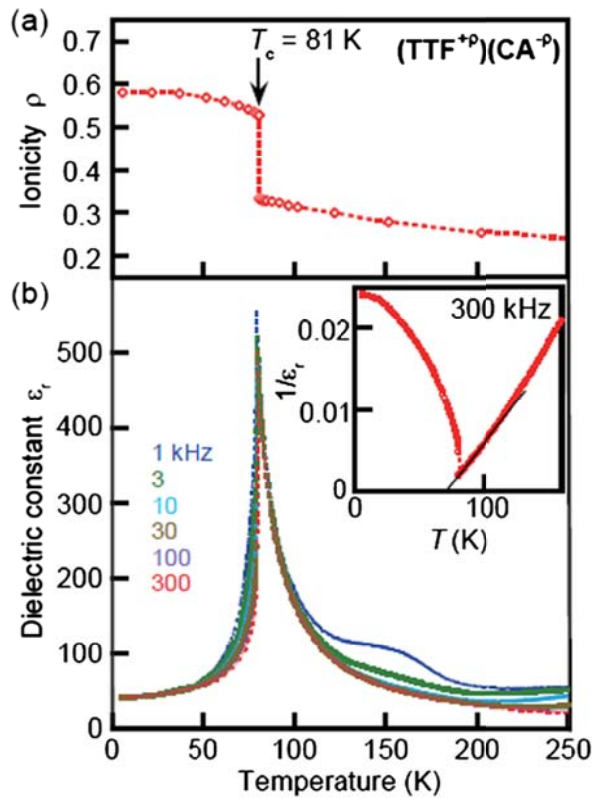

**Figure 10.** (a) Temperature-dependent ionicity of TTF-CA signifying its neutral-to-ionic transition [6]. (b) Temperature dependence of the dielectric constant for various frequencies. The inset shows the inverse dielectric constant at 300 kHz [6]. (Reprinted with permision from [6]. Copyright 2012 by the American Physical Society.)

*4.2 Two-dimensional charge-transfer salts*

*4.2.1 The relaxor ferroelectrics κ-(ET)₂Cu₂(CN)₃ and β'-(ET)₂ICl₂*

During recent years, dielectric spectroscopy on a number of two-dimensional charge-transfer salts has revealed the typical signatures of relaxor ferroelectricity [8,9,10,120]. The most prominent example is κ-(ET)$_2$Cu$_2$(CN)$_3$ [8]. Here ET stands for bis(ethylenedithio)-tetrathiafulvalene (often also abbreviated as BEDT-TTF). It consists of alternating conducting ET and insulating Cu$_2$(CN)$_3$ layers, stacked along the *a* axis. Figure 11 schematically indicates the structure within the ET planes. Pairs of molecules (thick grey lines) are dimerized and share a single hole, indicated by the red spheres; the shaded areas indicate their delocalization on the dimer. Reflecting the resulting Mott insulator phase, the system is termed a "dimer Mott insulator" [8,65,121]. The dimers, and thus the holes, are located on a triangular lattice (figure 11). Thus, the antiferromagnetic interaction of the hole spins leads to frustration and the material is a spin-liquid [122]. The resistivity of κ-(ET)$_2$Cu$_2$(CN)$_3$ exhibits no anomalies and no long-range CO seems to form in this material. Its temperature dependence within the planes follows VRH behaviour [equation (3)] with an exponent $\gamma$ of 1/3 indicating two-dimensional hopping conductivity [65].

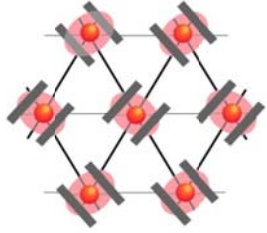

**Figure 11.** Schematic representation of the ET layers in κ-(ET)$_2$Cu$_2$(CN)$_3$ at high temperatures [7]. The thick grey lines represent the ET molecules. The red spheres denote the average positions of the holes at the centre of the dimer; the red shaded areas symbolize their delocalization.

The dielectric response of κ-(ET)$_2$Cu$_2$(CN)$_3$ was investigated by Abdel-Jawad et al. [8] and revealed clear evidence for relaxor ferroelectric behaviour. This is shown in figure 12(a), presenting the dielectric constant vs. temperature for various frequencies. This measurement was done with the electrical field directed out-of-plane. In many of the two-dimensional CT transfer salts this is the only way to achieve reasonable dielectric results as the in-plane conductivity often is too high to enable reliable dielectric measurements. The behaviour revealed by figure 12(a) is typical for relaxor ferroelectrics [cf. figure 3(c)]. In ref. [8], the frequency-induced shift of the peak temperatures in figure 12(a) were shown to be describable by the VFT law, equation (2) ($T_{VF} \approx 6$ K), again typical for relaxor ferroelectrics (see section 3.3). The envelope curve obtained by combining the high-temperature flanks of the $\varepsilon'(T)$ peaks for all frequencies represents the static dielectric constant. It was reported to follow the Curie-Weiss law [section 3.3, eq. (5)] below 60 K [8]. In this context, it should be noted that in relaxors usually strong deviations from Curie-Weiss behaviour show up and only at very high temperatures a linear $1/\varepsilon'$ vs. *T* behaviour is observed [85]. Nevertheless, fits by this law may provide at least a rough estimate of the quasistatic freezing temperature. $T_{CW} \approx 6$ K, obtained in this way agrees with $T_{VF}$ in this system.

In [8], the found relaxor ferroelectricity of κ-(ET)$_2$Cu$_2$(CN)$_3$ was qualitatively explained considering fluctuations of the holes between the two molecules forming a dimer. These fluctuations correspond to the reorientation of a dipole which leads to the observed relaxational response. Thus this materials seems to represent an electronic relaxor ferroelectric. To explain the relaxor behaviour, it was suggested that, due to the zigzag pattern of the ET molecules, the dipole-dipole interactions are relatively weak and compete with the intra-dimer fluctuations, preventing the formation of a conventional long-range ordered ferroelectric state [8]. Subsequently, several theoretical works have been published explaining the interesting dielectric behaviour of κ-(ET)$_2$Cu$_2$(CN)$_3$ [123,124,125], essentially following the physical picture proposed in ref. [8] assuming dipoles created by charge disproportionation on molecular dimers. This material is a spin liquid, i.e., the magnetic degrees of freedom are disordered. The polar degrees of freedom in relaxors in many respects also exhibit the characteristics of disordered systems. It is suggestive to see a connection here, especially when considering the results on $\kappa$-(ET)$_2$Cu[N(CN)$_2$]Cl (see section 4.2.3), which has the same pattern of the ET molecules but exhibits both magnetic and polar order, occurring at nearly the same temperature. Indeed in ref. [124] electric dipoles on dimers are considered to significantly modify the spin-exchange coupling, resulting in a "dipolar-spin liquid".

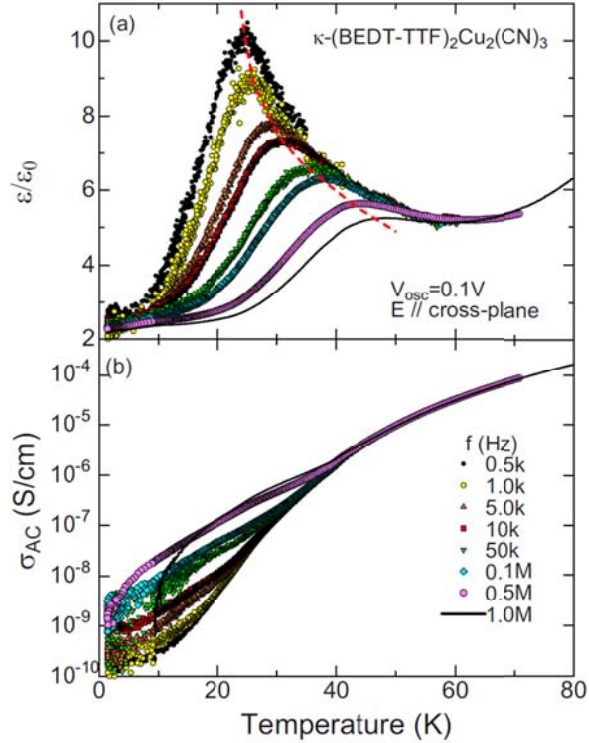

**Figure 12.** (a) Temperature dependence of the dielectric constant (a) and the conductivity (b) of $\kappa$-(ET)$_2$Cu$_2$(CN)$_3$ as measured at various frequencies [8]. The dotted line in (a) indicates the peak temperatures. (Reprinted with permision from [8]. Copyright 2012 by the American Physical Society.)

Very recently, the dielectric results of ref. [8] where nicely corroborated by dielectric investigations of three different crystals of $\kappa$-(ET)$_2$Cu$_2$(CN)$_3$ [65]. In contrast to [8], measurements were performed with the electric field oriented in all three crystallograpic directions. As noted by the authors of ref. [65], relaxational behaviour was found for all field orientations, revealing the "typical fingerprints of relaxor ferroelectricity". In this work, the frequency dependence of the dielectric constant and loss was evaluated and fitted by the CC function [eq. (1) with $\beta_{HN} = 1$]. Thus, it should be noted that the obtained relaxation time cannot be directly compared to $\tau$ reported in ref [8], where it was deduced from the real part. A significant sample dependence of the dielectric properties was revealed by the performed measurements, which seems to be a rather common problem of the organic charge transfer salts. The temperature dependence of the relaxation times obtained from the CC fits performed in [65] partly followed Arrhenius behaviour. In one sample, interestingly it was found to level off at low temperatures becoming nearly temperature independent. As noted by the authors, this indicates a transition to tunneling-dominated dynamics at low temperatures. In that work, in agreement with [8], a charge disproportionation $\delta$ within the ET dimers of about $0.1e$ was estimated from the found Curie-Weiss behaviour, eq. (5). (However, it should be noted that, based on the short-range clusterlike polar order assumed for relaxors [84,85], this should not lead to long-range CO.) The authors of [65] pointed out that this value of $\delta$ in not in accord with recent results from optical measurements [126] and suggested "charge defects generated in interfaces" between domains to be responsible for the observed effects in $\kappa$-(ET)$_2$Cu$_2$(CN)$_3$.

The conductivity of $\kappa$-(ET)$_2$Cu$_2$(CN)$_3$, shown in figure 12(b) is not discussed in detail in ref. [8]. Neglecting obvious experimental artefacts at the two highest frequencies, which lead to a strong additional decline of $\sigma(T)$ at low temperatures, a dc and an ac response can be discerned [cf. equation (4a)]. The dc conductivity corresponds to the curve at 0.5 kHz from which the other curves deviate at different temperatures, depending on temperature. Weak shoulders are superimposed to the general decrease of $\sigma(T)$ with decreasing temperature (occurring, e.g., somewhat below 30 K for the curve at 1 MHz). They probably correspond to the relaxational behaviour seen in $\varepsilon'(T)$. The crossover to a weaker temperature dependence of the conductivity, observed, e.g., below about 40 K for 1 MHz, indicates ac conductivity due to hopping charge transport (section 3.2).

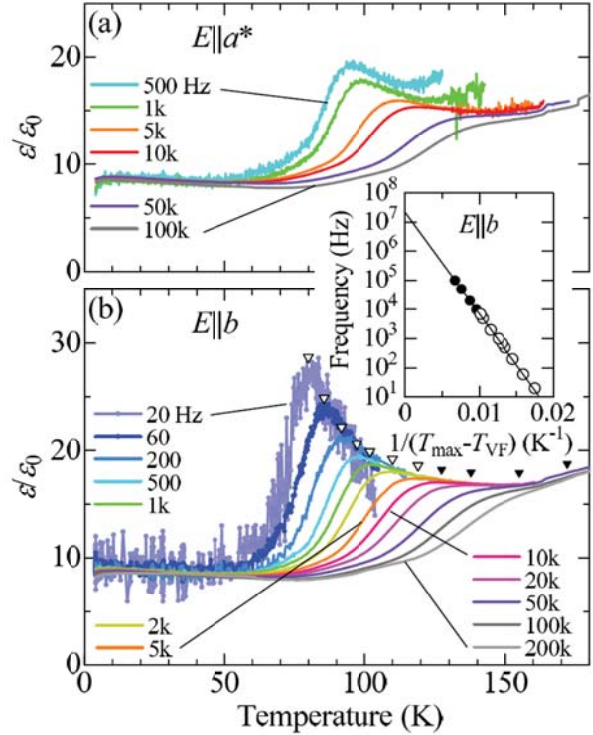

**Figure 13.** Temperature-dependent dielectric constant of $\beta'$-(ET)$_2$ICl$_2$ for various frequencies [9]. Results for electrical-field directions perpendicular (a) and parallel (b) to the ET planes are shown. The inset demonstrates VFT behaviour of the peak frequency. (Reprinted with permission from [9]. Copyright 2013 by the American Physical Society.)

Recently, relaxor ferroelectricity has also been detected in $\beta'$-(ET)$_2$ICl$_2$ [9]. In contrast to the spin-liquid state of $\kappa$-(ET)$_2$Cu$_2$(CN)$_3$, it exhibits antiferromagnetic ordering at $T_N \approx 22$ K [127], i.e. in this material long-range order of the spins but only short-range, clusterlike order of the dipolar degrees of freedom evolves. In $\beta'$-(ET)$_2$ICl$_2$, the ET molecules also are dimerized but are arranged in a square lattice within the $bc$ planes instead of the triangular lattice of $\kappa$-(ET)$_2$Cu$_2$(CN)$_3$. Just as for the latter, the dipolar degrees of freedom were assumed to arise from charge disproportionation within the dimers [9]. As shown in figure 13 [9], the temperature dependence of the dielectric constant reveals the typical features of relaxor ferroelectricity. This is the case for both in plane and out-of-plane field directions [figures 13(a) and (b), respectively]. As for $\kappa$-(ET)$_2$Cu$_2$(CN)$_3$, Curie-Weiss behaviour is found at the right flanks of the $\varepsilon'(T)$ peaks ($T_{CW} \approx 67$ K). Pyrocurrent measurements have revealed an onset of polarization at a similar temperature of about 60 K [9]. It should, however, be noted that generally in relaxor ferroelectrics sizable polarization occurs rather far above the $\varepsilon'(T)$ peak temperatures [85]. The inset of figure 13 demonstrates VFT temperature dependence of the $\varepsilon'(T)$ peak temperatures ($E\|b$) with $T_{VF} \approx 23$ K. Thus, in $\beta'$-(ET)$_2$ICl$_2$ both characteristic temperatures do not agree, i.e. $T_{CW} \neq T_{VF}$. In principle, there is no general rule that both temperatures should agree in relaxor ferroelectrics, especially as in most relaxors the Curie-Weiss law is applicable far above the peak temperature only [85] and the VFT law is a purely phenomenological parameterization. Nevertheless, this difference for $\beta'$-(ET)$_2$ICl$_2$, in contrast to the findings in $\kappa$-(ET)$_2$Cu$_2$(CN)$_3$, seems interesting and it also should be noted that $T_{VF}$ and $T_N$ are nearly identical. In ref. [9] "frustration between ferroelectric and antiferroelectric interactions" and "spin-charge coupled degrees of freedom" were considered to rationalize the finding of $T_{CW} > T_{VF}$.

4.2.2 $\alpha$-(ET)$_2$I$_3$, a relaxor ferroelectric?

Recently, second-harmonic generation (SHG) measurements of $\alpha$-(ET)$_2$I$_3$ suggested possible ferroelectric ordering in this material [128,129]. SHG points to a non-centrosymmetric crystal structure [130,131]. However, it should be noted that this does not necessarily imply ferroelectricity. Ferroelectrics, in addition also are characterized by a unique polar axis and switchable polarization [77]. Only 10 of the 21 non-centrosymmetric crystal structures are polar and only a fraction of the materials belonging to these groups are ferroelectric [77]. Thus, dielectric and polarization measurements are necessary to clearly prove ferroelectricity in $\alpha$-(ET)$_2$I$_3$. Again, in this two-dimensional material the ET molecules are located in planes and have an average charge of +0.5 per molecule. The $\alpha$ structure in the ET planes corresponds to molecules that are arranged in a herringbone pattern (figure 14) [132]. Interestingly, only part of the molecules are dimerized and arranged in stacks (denoted by I in figure 14) that are separated by stacks of non-dimerized molecules (stacks II). This material is known to show a clear CO transition at $T_{CO} \approx 135$ K [132,133]. It is accompanied by a huge drop of the in-plane conductivity as was shown in two pioneering works by Bender *et al*. [132,134]. Later on, this finding was confirmed in refs. [135] and [136].

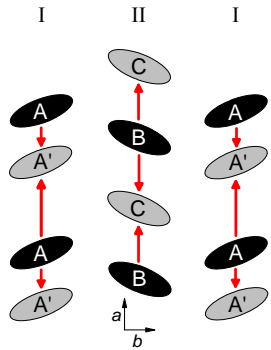

**Figure 14.** Schematic plot of an ET plane of $\alpha$-(ET)$_2$I$_3$ visualizing the dimerization (strongly exaggerated in the plot) and CO pattern [10]. Molecules with higher charge values are shown in black. Dimerization only exists in stack I. The dipolar moments, adding up to a net polarization in the dimerized stacks, are indicated by arrows [10].

In ref. [135], in addition to the dc response, also the in-plane permittivity and ac conductivity of $\alpha$-(ET)$_2$I$_3$ were reported in a wide range, including microwave to optical frequencies. From spectra of the dielectric constant shown in that work, in the kHz range and at 60 - 120 K a relaxation-like mode with huge amplitude was detected with the static dielectric constant reaching values up to $5 \times 10^5$. In two more recent papers [11,136], the existence of this huge relaxation mode was confirmed and at low temperature (47 K) even a second relaxation was reported as shown in figure 15, which astonishingly shows no significant temperature variation of its relaxation time. The larger and slower mode was ascribed to "the phasonlike excitation of the $2k_F$ bond-charge density wave" while the faster relaxation process was assumed to arise from the "motion of domain wall pairs" within the CO structure [11]. In these measurements the in-plane dielectric response was detected. Within the ET planes, even in the charge-ordered phase the conductivity of $\alpha$-(ET)$_2$I$_3$ is relatively high [136]. Its frequency dependence was found to follow a power-law pointing to hopping transport (section 3.2) [136]. Generally, dielectric measurements of conducting materials are difficult. While the authors of ref. [11,136] succeeded in this demanding task, the high conductivity and/or the observed CDW-like, very strong relaxation mode with colossal values of $\varepsilon_s$ may obscure the signatures of ferroelectric order in the dielectric properties, expected on the basis of the mentioned SHG measurements [128,129]. Moreover, to arrive at the loss peaks shown in Fig. 15, the authors had to subtract the dc conductivity, which may lead to some ambiguities and large data scattering as discussed, e.g., in ref. [137]. As mentioned above, a solution of these problems may be out-of-plane dielectric measurements, where the conductivity can be expected to be significantly lower. While such measurements were done by the authors of [136], which they reported to have revealed similar behaviour as for in-plane field direction, unfortunately no spectra were provided. It is clear that for $\alpha$-(ET)$_2$I$_3$, the ferroelectric polarization should be predominantly oriented parallel to the ET planes but in ref. [10] arguments for an out-of-plane component of the polarization were provided, based on an asymmetric charge distribution along the long axis of the ET molecules.

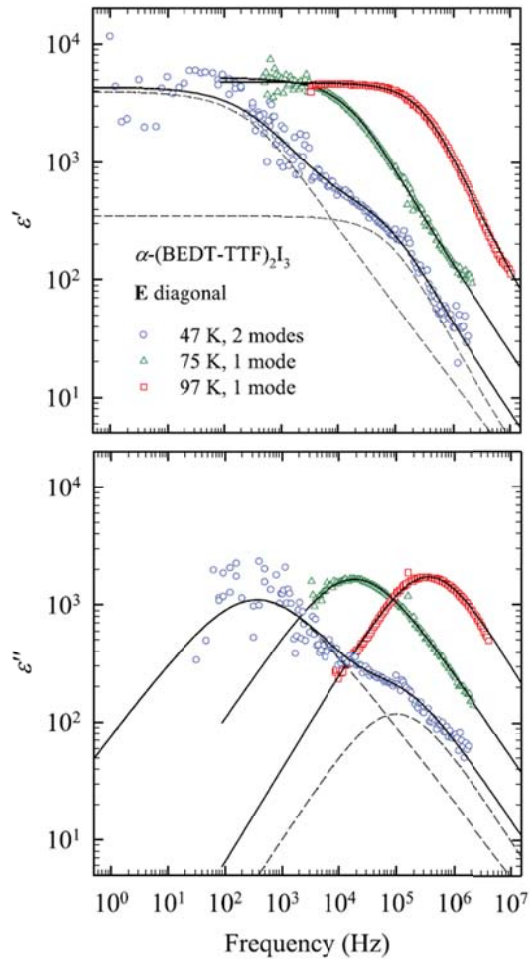

**Figure 15.** Frequency dependence of the in-plane dielectric constant and loss of $\alpha$-(ET)$_2$I$_3$ at three temperatures [136]. The lines are fits with a single CC function (75 and 97 K) or the sum of two CC functions (47 K). (Reprinted with permision from [136]. Copyright 2011 by the American Physical Society.)

In ref. [24], out-of-plane measurements were reported but, unfortunately, only results below 50 K were provided. The dielectric-constant spectra of that work are shown in figure 16, including also results for in-plane field direction, which we discuss first. The investigated frequency range partly overlaps with that covered in refs. [11,136] (figure 15). In agreement with those works, $\varepsilon'(\nu)$ within the ET planes reaches very large values at low frequencies. However, when comparing the spectra at 47 K in figure 15 and at 50 K in figure 16, clear discrepancies concerning the absolute values of $\varepsilon'$ are revealed. Moreover, at low frequencies all curves in figure 15 seem to approach a static dielectric constant of about $5 \times 10^3$, typical for relaxational response (section 3.1). In contrast, in figure 16 no saturation at low frequencies is observed. Instead, $\varepsilon'(\nu)$ in figure 16 approaches a plateau at high frequencies with values of several 100, of which there is no trace in figure 15, despite the spectra extend to much higher frequencies. These marked discrepancies may indicate a strong sample-to-sample variation but they may also signify the mentioned difficulties of dielectric measurements for samples that have rather high conductivity. In ref. [24] the temperature dependence of the in-plane dielectric constant was ascribed to the polarization of electron-hole pairs.

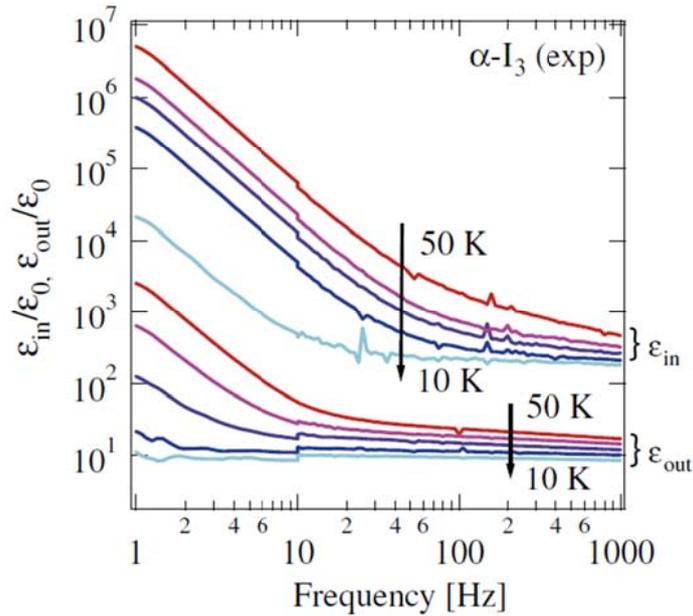

**Figure 16.** Frequency dependence of the dielectric constant of $\alpha$-(ET)$_2$I$_3$ at five temperatures (10 K steps) for in ($\varepsilon_{in}$) and out-of-plane ($\varepsilon_{out}$) field directions [24]. (Reprinted from [24] with kind permision from the Physical Society of Japan.)

The dielectric constant for out-of-plane field direction, reported in ref. [24], generally is much lower than for the in-plane measurements (figure 16). The spectra approach a reasonable high-frequency limit of about 10. At low frequencies, values up to several 1000 are reached. Unfortunately, a ferroelectric origin of these rather high values of $\varepsilon'$ cannot be deduced from these measurements as no indications for a peak in $\varepsilon'(T)$ are found. However, this can be ascribed to the restricted temperature range of these investigations performed at $T \leq 50$ K only.

Very recently, a dielectric investigation of $\alpha$-(ET)$_2$I$_3$ for out-of-plane field direction was performed by our group in collaboration with scientists from the Goethe-University Frankfurt, covering a broader temperature range from 4 - 300 K and frequencies from 1 Hz to several MHz [10]. Figure 17 presents the obtained conductivity at 1 Hz, which is virtually identical to the dc conductivity [10]. For comparison, also the in-plane conductivity results from ref. [136] are shown. While there is already a small anisotropy within the plane as discussed in ref. [136], as expected the out-of-plane conductivity is markedly lower and differs from the in-plane values by about 3 - 4 orders of magnitude. Interestingly, all three curves can rather well be scaled onto each other (see inset of figure 17), including the strong reduction of $\sigma$ at the CO transition at 135 K. This implies that the reduction of the number of free charge carriers at $T_{CO}$ affects both the in-plane and out-of-plane conductivity in the same way.

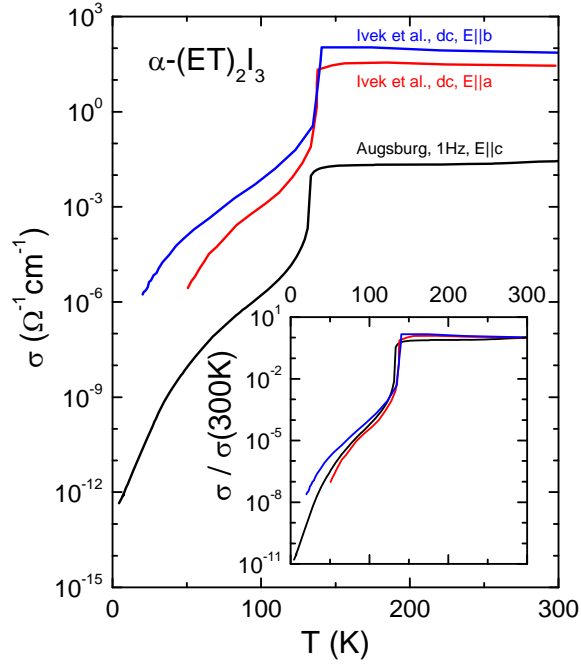

**Figure 17.** Temperature dependence of the dc conductivity of $\alpha$-$(ET)_2I_3$ for the electrical field directed within (data taken from ref. [136]) and perpendicular to the ET planes [10]. Inset: Scaled temperature dependences of the curves shown in the main frame.

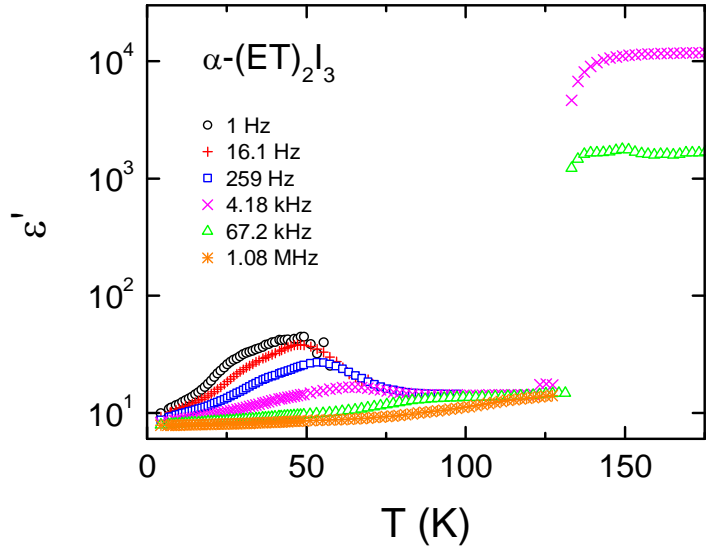

**Figure 18.** Temperature dependence of the dielectric constant of $\alpha$-$(ET)_2I_3$ for various frequencies [10].

The much lower out-of-plane conductivity of $\alpha$-$(ET)_2I_3$, revealed in figure 17, enabled reliable measurements of the dielectric permittivity in the charge ordered state [10]. As shown in figure 18, even some results above $T_{CO}$ could be obtained. Just as for the conductivity, the CO transition is clearly revealed in $\varepsilon'(T)$, which seems to increase by about 2 decades when leaving the CO state. Nevertheless, it should be noted that the significance of the results at $T > T_{CO}$ is limited as here the conductivity still is too high for the measurement device to produce reliable data and because Maxwell-Wagner effects (section 3.4) may start to play a role. Interestingly, immediately below $T_{CO}$, where no such problems exist, $\varepsilon'(T)$ is not or only weakly temperature dependent (depending of frequency). The SHG signal detected in ref. [128] was reported to continuously arise below $T_{CO}$. Therefore, if assuming that the SHG is caused by ferroelectric ordering, the ferroelectric transition in $\alpha$-$(ET)_2I_3$ would be expected to be identical with $T_{CO}$ where an anomaly in $\varepsilon'(T)$ should occur. While no clear statement at $T > T_{CO}$ can be made, due to the mentioned experimental problems, at $T < T_{CO}$ clearly the absence of any anomalous behaviour in $\varepsilon'(T)$ that would point to a ferroelectric transition at $T_{CO}$ can be stated. Instead, unexpectedly a broad peak in $\varepsilon'(T)$ shows up at much lower temperatures, around 50 K. Its frequency and temperature dependence (figure 18 and ref. [10]) bears the clear signatures of relaxor ferroelectrics (section 3.3.3). This finding is unexpected and not fully explained until now. In [10], we have suggested that, just as proposed for $\kappa$-(ET)$_2$Cu$_2$(CN)$_3$, $\beta'$-(ET)$_2$ICl$_2$, and $\kappa$-(ET)$_2$Cu[N(CN)$_2$]Cl, the ordering of electronic dipolar degrees of freedom leads to the ferroelectricity in $\alpha$-(ET)$_2$I$_3$. The relaxor behaviour and decoupling of the ferroelectric ordering from the CO transition was ascribed to the peculiarities of the $\alpha$ structure of the molecules within the ET planes. Especially it was argued that the alteration of dimerized and undimerized stacks may prevent the formation of canonical long-range ferroelectric order directly below $T_{CO}$. Notably, in contrast to canonical relaxor behaviour [84,85], the static dielectric constant tends to saturate at a finite value for high temperatures (figure 18), for which an explanation was proposed in ref. [10] in terms of the scenario discussed above. Interestingly, similar behaviour is also found in the relaxors $\kappa$-(ET)$_2$Cu$_2$(CN)$_3$ (figure 12 [8]) and $\beta'$-(ET)$_2$ICl$_2$ (figure 13 [9]).

The frequency dependence of the dielectric permittivity (real and imaginary part) shown in [10] also signifies relaxational response and could be well fitted by a CC function (eq. 1 with $\beta_{HN} = 1$), including an additional dc-conductivity contribution $i\sigma_{dc} / (2\pi\nu\varepsilon_0)$. Finally, polarization experiments that could be successfully performed at low temperatures, far within the insulated range, also revealed clear evidence for ferroelectricity in $\alpha$-(ET)$_2$I$_3$. The out-of-plane $\varepsilon'$ spectra from ref. [24], shown in figure 16, qualitatively agree with the relaxor behaviour detected in [10]. Especially, $\varepsilon'$ continuously increases with temperature for the investigated range 10 K $< T <$ 50 K. Moreover, the frequency dependence may be consistent with relaxor ferroelectricity but the absolute values reached at low frequencies are clearly higher than in [10] pointing to sample-dependent differences or non-intrinsic effects.

### 4.2.3 Multiferroic $\kappa$-(ET)$_2$Cu[N(CN)$_2$]Cl

$\kappa$-(ET)$_2$Cu[N(CN)$_2$]Cl, is an especially interesting case, as recently dielectric and polarization experiments were reported revealing strong hints at ferroelectricity that occurs simultaneously with magnetic order [7]. This finding characterizes this material as multiferroic, making it a member of one of the most prominent material classes in current materials science [18,19,20]. Moreover, alternative explanations were proposed to explain the remarkable dielectric properties of $\kappa$-(ET)$_2$Cu[N(CN)$_2$]Cl [7,22,137,138,139] and our microscopic understanding of this material still is far from complete. The structure of the ET planes in this compound is very similar as for $\kappa$-(ET)$_2$Cu$_2$(CN)$_3$ shown in figure 11. However, as was pointed out in ref. [139] providing a detailed comparison of both systems, the dimer centres in the Cl system are arranged in more anisotropic triangles. This may lead to reduced frustration and could explain the fact that this material shows antiferromagnetic ordering at about 25 - 30 K instead of a spin-liquid state as in $\kappa$-(ET)$_2$Cu$_2$(CN)$_3$ [140,141] (see below for an alternative explanation).

In the following, we first treat the dc transport of $\kappa$-(ET)$_2$Cu[N(CN)$_2$]Cl, which is of relevance for the interpretation of its dielectric behaviour. There are several works where marked anomalies in the temperature dependence of the conductivity or resistivity of $\kappa$-(ET)$_2$Cu[N(CN)$_2$]Cl were reported. An early example can be found in ref. [142] where especially for the out-of-plane field direction a strong anomalous low-temperature increase of $\rho(T)$ was detected at about 30 K, not far from the magnetic ordering temperature $T_N$. $\sigma(T)$ calculated from $\rho(T)$ shown in this work is included in figure 19a (solid line). As a second example, results from ref. [143] are included (dashed line). Finally, figure 19a contains the 2.1 Hz curve from ref. [7], which, based on the frequency independence of $\sigma'$ at low frequencies (inset), is a good approximation of the dc conductivity. While the absolute values partly are different, all three curves show the mentioned strong reduction of the conductivity close to $T_N$, which is indicated by the arrow. A similar strong conductivity anomaly around 30 K also was reported in ref. [144]. As CO is a common phenomenon in related systems, some of them being treated in the preceding sections, one may speculate that these anomalies indicate a CO transition in $\kappa$-(ET)$_2$Cu[N(CN)$_2$]Cl, leading to a reduction of $\sigma$ similar as revealed in figures 6 and 17. However, it should be noted that in other works no such anomaly was found in $\kappa$- $(ET)_2Cu[N(CN)_2]Cl$ (e.g., [22,145]) and, obviously, there is a strong sample-to-sample variation of its conductivity behaviour, as also explicitly shown in [137].

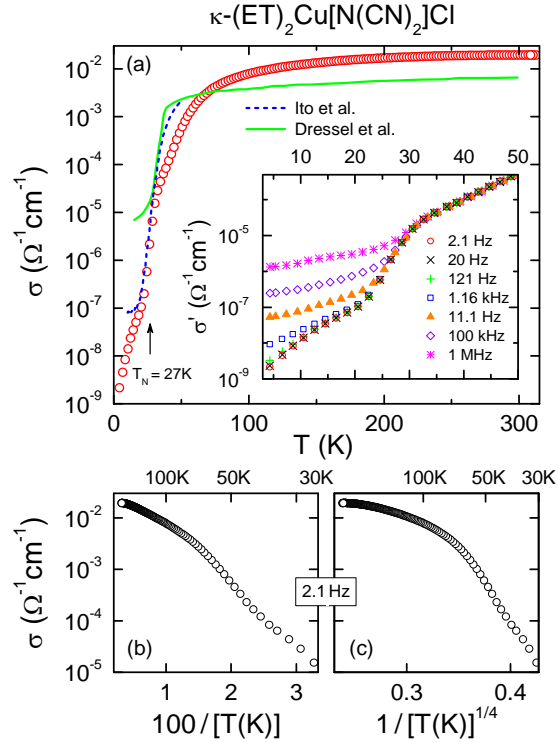

**Figure 19.** (a) Temperature dependence of the out-of-plane dc conductivity of $\kappa$-$(ET)_2Cu[N(CN)_2]Cl$ taken from refs. [7] (circles), [142] (solid line) and [143] (dashed line). The inset shows $\sigma'$ at different frequencies [7] demonstrating that the 2.1 Hz curve shown in the main frame corresponds to the dc conductivity. Frame (b) provides the conductivity at 2.1 Hz, plotted in an Arrhenius representation. In (c) the same data are shown in a representation that should lead to linear behaviour for the VRH $T^{1/4}$-law [7].

The frequency dependence of $\sigma'$, revealed at low frequencies and temperatures in the inset of figure 19(a), follows the UDR, eq. (4a), indicating hopping conductivity [7]. This is corroborated by figure 19(b) showing that $\sigma'(T)$ at 2.1 Hz does not follows a simple thermally activated behaviour in any extended temperature range. However, frame (c) demonstrates that it also cannot be described by Mott's VRH law, eq. (3), and thus the microscopic mechanism of the hopping process in $\kappa$-$(ET)_2Cu[N(CN)_2]Cl$ still remains to be clarified.

Concerning the dielectric properties of $\kappa$-$(ET)_2Cu[N(CN)_2]Cl$, we first want to mention the pioneering work by Pinterić *et al*. [138] who found a strongly frequency-dependent permittivity in two different single crystals and large values of $\varepsilon' > 1000$ at low frequencies. The obtained spectra were interpreted in terms of relaxational response. The spectra on crystal 2, shown in figure 20, were fitted by the sum of a Debye (slower mode) and a HN function [faster mode; eq. (1)] [138]. The two corresponding relaxation processes, both with relaxations strengths larger than 1000, were ascribed to "charged domain walls" and the "domain structure of the $Cu^{2+}$ subsystem". One should be aware that these data can also be fitted by a single distributed relaxation function, namely the CD function. Indeed, later data by the same group were parameterized in this way [22] as will be discussed below. It is clear that in these in-plane measurements the results are hampered to some extent by strong conductivity contributions, which were subtracted from the raw data to arrive at the shown loss spectra. As demonstrated for $\kappa$-$(ET)_2Cu[N(CN)_2]Cl$ in ref. [137], the occurrence of peaks in data corrected in this way critically depends on the exact choice of the subtracted dc conductivity, introducing some ambiguities in their interpretation. In any case, the results of ref. [137] already indicated an interesting dielectric response of $\kappa$-$(ET)_2Cu[N(CN)_2]Cl$ and especially the found very high values of $\varepsilon'$, if of intrinsic nature, made this material attractive for further investigations. Unfortunately, the dielectric measurements of ref. [137] were restricted to temperatures below 30 K (probably due to the too high conductivity at higher temperatures) and no temperature-dependent plot of $\varepsilon'$ was provided, which could reveal the signature of possible ferroelectric of relaxor-ferroelectric behaviour as found in the related systems treated above.

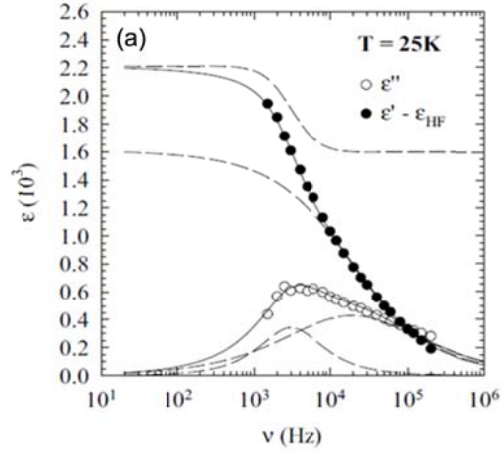

**Figure 20.** (Frequency dependence of the real and imaginary parts of the dielectric constant of $\kappa$-(ET)$_2$Cu[N(CN)$_2$]Cl at 25 K for in-plane field direction (taken from [138] with kind permission from Springer Science and Business Media). The solid lines show fits with two relaxation functions, whose individual contributions are indicated by the dashed lines.

These problems were overcome in ref. [7], where a detailed investigation of $\kappa$-(ET)$_2$Cu[N(CN)$_2$]Cl was provided for the electrical field directed perpendicular to the ET planes. The much lower out-of-plane conductivity reported in [7] enabled the unequivocal detection of the dielectric constant up to about 50 K, extending beyond the temperature of the mentioned conductivity anomaly occurring close to the magnetic ordering temperature $T_N$. Figure 21 provides the most important results of this work, namely the occurrence of peaks in the temperature dependence of the dielectric constant, arising somewhat below $T_N$. For increasing frequency, the amplitude of these peaks becomes reduced and virtually no anomaly is found for the highest frequency of 1 MHz, but their positions exhibit no or only weak temperature shifts, especially at low frequencies. As pointed out in [7], the overall signatures of $\varepsilon'(T)$ revealed in figure 21 closely resemble those of typical order-disorder ferroelectrics as discussed in section 3.3.2 [cf. figure 3(b)]. This was nicely corroborated by polarization measurements revealing, e.g., the typical $P(E)$ hysteresis loops expected for ferroelectrics [7]. It should be noted that the continuous decrease of $\varepsilon'(\nu)$ found in ref. [138] (figure 20) at least qualitatively agrees with the results of figure 21, despite the absolute values at low-frequencies are higher, which is not unexpected if considering the different field directions of the experiments and some sample dependence of the dielectric properties. In agreement with the spectra reported in [138], no indication of a low-frequency saturation of $\varepsilon'(\nu)$ was found for the results of ref. [7] as explicitly shown in ref. [137]. Moreover, as mentioned above, the appearance of a peak in $\varepsilon''(\nu)$ was demonstrated to critically depend on very slight variation of the subtracted dc value [137]. Therefore, based on the dielectric results of refs. [7] and [138], there is no clear evidence of relaxational response in $\kappa$-(ET)$_2$Cu[N(CN)$_2$]Cl. To reveal a possible relaxation process that may be expected for order-disorder ferroelectrics (section 3.3.2), measurements at lower frequencies may be necessary. The combined right flanks of the $\varepsilon'$ peaks in figure 21 could be described by a Curie-Weiss law, eq. (5) with an additive background dielectric constant of about 120, mainly ascribed to stray capacitance [7]. Interestingly, the obtained Curie-Weiss temperature $T_{CW} \approx 25$ K is close to the $T_N$ values reported for this material [137,140,141].

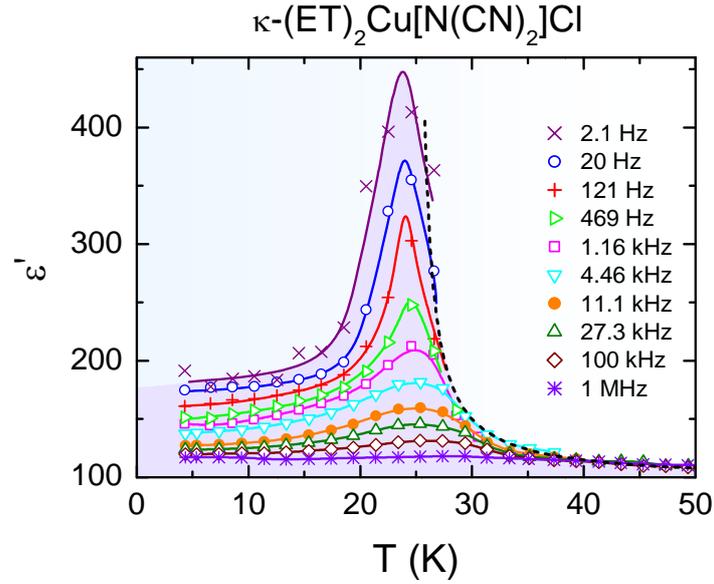

**Figure 21.** Temperature dependence of the out-of-plane dielectric constant of $\kappa$-(ET)$_2$Cu[N(CN)$_2$]Cl, measured at various frequencies (reprinted from [7]). The solid lines are guides for the eyes. The dashed line indicates Curie-Weiss behaviour ($T_{CW}$ = 25 K).

In ref. [7], even for the in-plane field direction a peak in $\varepsilon'(T)$ was reported at a similar temperature, signifying ferroelectric ordering. However, it is partly superimposed by a second peak arising from a relaxation process, which most likely is non-intrinsic and of Maxwell-Wagner type (section 3.4). This finding is consistent with $\varepsilon'(T)$ curves reported in ref. [146] (figure 22), which were restricted to temperatures below 25 K and a relatively narrow frequency range. In figure 22, even the onset of a peak is observed close to the upper temperature limit of 25 K of this investigation, in good qualitative agreement with the results in [7]. In [146] the dielectric results and the found pronounced nonlinear IV characteristics were explained in terms of "bound pairs of an electron and a hole" that are "thermally excited in a charge-ordered state".

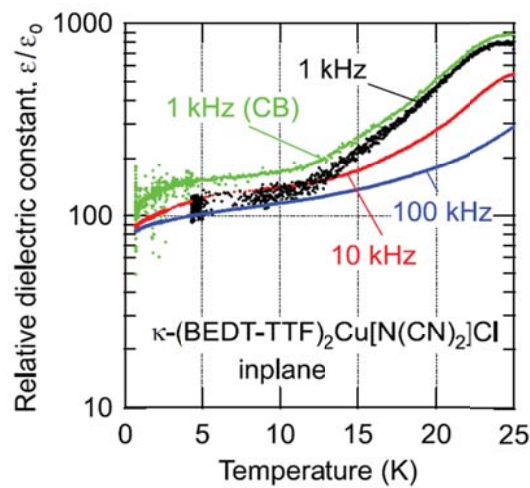

**Figure 22.** Temperature dependence of the in-plane dielectric constant of $\kappa$-(ET)$_2$Cu[N(CN)$_2$]Cl at various frequencies (taken from [146]). The curve denoted by "CB" was measured with a different device. (Reprinted with permision from [146]. Copyright 2011 by the American Physical Society.)

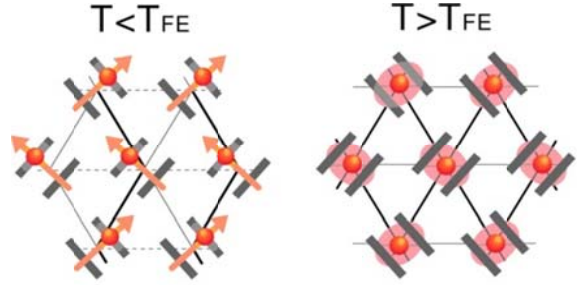

**Figure 23.** Schematic structure within the ET planes of $\kappa$-(ET)$_2$Cu[N(CN)$_2$]Cl as suggested in ref. [7]. The grey bars represent the ET molecules, forming dimers that are located on a triangular lattice indicated by thin lines. In the left part of the figure the situation in the suggested ferroelectric state is shown with the holes (spheres) locked-in to one of the two ET molecules forming a dimer. The arrows indicate the resulting dipolar moment. The right part illustrates the structure above the ferroelectric transition, where the holes are delocalized within the dimers, indicated by the shaded areas.

The finding of ferroelectricity suggested in ref. [7] and the magnetic order below $T_N \approx T_{CW}$ indicates multiferroicity and a close coupling of dipolar and magnetic degrees of freedom. There are various mechanisms of multiferroicity, the most prominent one being the spin-driven ferroelectricity arising via the Dzyaloshinskii-Moriya interaction, mentioned in section 3.3.4 [147]. However, as discussed in [7,137], the absence of any influence of external magnetic fields up to 9 T on the dielectric properties makes such a scenario unlikely for $\kappa$-(ET)$_2$Cu[N(CN)$_2$]Cl and, thus, in ref. [7] a different mechanism was proposed. The essential idea is illustrated in figure 23. At high temperatures, $T > T_N \approx T_{CW}$, a similar situation as for $\kappa$-(ET)$_2$Cu$_2$(CN)$_3$ (figure 11) can be assumed, where the holes shared by two neighbouring ET molecules are delocalized on the dimers (right part of figure 23). As the magnetism arises from the spins of the holes, the triangular lattice leads to geometrical frustration and no spin order is formed. At low temperatures, CO within the dimers occurs (left part of figure 23) leading to macroscopic polarization and ferroelectricity. Thus within this scenario the dipolar degrees of freedom in this system have similar origin as suggested for the relaxors $\kappa$-(ET)$_2$Cu$_2$(CN)$_3$ and $\beta'$-(ET)$_2$ICl$_2$ (see discussion of these systems in section 4.2.1). CO occurring close to $T_N \approx T_{CW}$ is fully consistent with the conductivity anomaly documented in figure 19. As, due to the charge ordering, at low temperatures the spins no longer are located on a regular triangular lattice (left part of figure 23), frustration is broken and spin order is triggered, occurring simultaneously with the ferroelectric order. This implies that in $\kappa$-(ET)$_2$Cu[N(CN)$_2$]Cl the spin order is driven by the ferroelectricity, in marked contrast to the spin-driven ferroelectricity in helical magnets and thus representing a new mechanism of magnetoelectric coupling.

Interestingly, in ref. [22] recently an alternative explanation of the dielectric behaviour of $\kappa$-(ET)$_2$Cu[N(CN)$_2$]Cl was considered. This was triggered by the fact that until now no evidence of charge disproportionation (implying CO), e.g., by FIR measurements, was found in this system [22,126]. The authors of ref. [22] suggested "short-range discommensurations of the commensurate antiferromagnetic phase" at 30 - 50 K and "charged domain-wall relaxations in the weak ferromagnetic state" at lower temperatures to account for the occurrence of the anomalous dielectric response of $\kappa$-(ET)$_2$Cu[N(CN)$_2$]Cl. Their conclusions partly are based on dielectric results interpreted in terms of a relaxation process similar to those in ref. [138] (figure 24). These data extend to lower frequencies than those in [138] from the same group (cf. figure 20) and were fitted with a single CD function [eq. (1) with $\alpha_{HN} = 0$]. Unfortunately, the data scatter of $\varepsilon'$ at low frequencies is too high to provide unequivocal evidence of relaxational behaviour, showing no significant approach of a plateau that would correspond to the static dielectric constant [cf. figure1(b)]. Moreover, the imaginary part is hampered by the mentioned subtraction effects of the dc response [137], which also leads to strong data scatter. Overall, the occurrence of a relaxational process in the frequency dependence of the permittivity of $\kappa$-(ET)$_2$Cu[N(CN)$_2$]Cl is an interesting question that remains to be clarified. In any case, it would be consistent with both scenarios as for order-disorder ferroelectrics relaxational response is also known to occur (section 3.3.2).

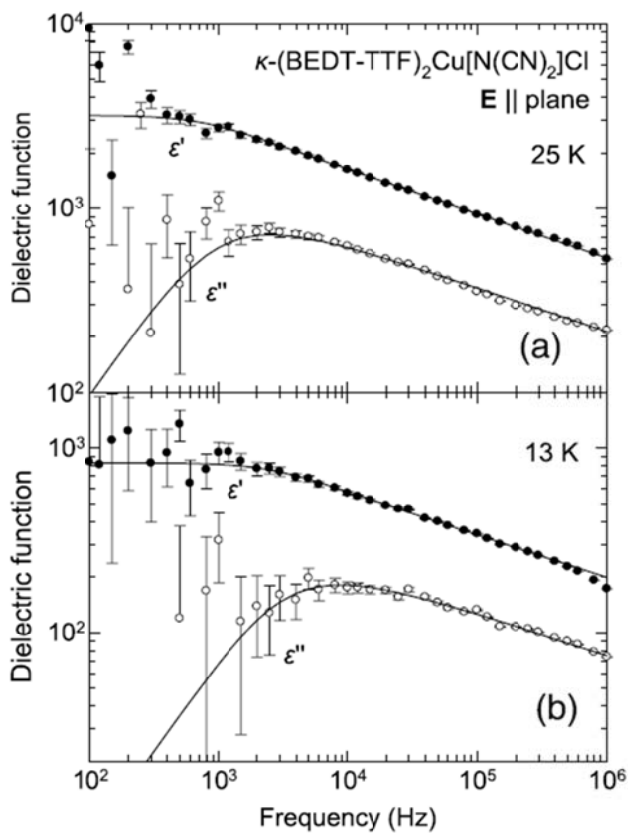

**Figure 24.** Frequency dependence of the in-pane dielectric permittivity of $\kappa$-(ET)$_2$Cu[N(CN)$_2$]Cl at two temperatures [22]. The lines are fits with the CD function. (Reprinted from [22]. © IOP Publishing. Reproduced by permission of IOP Publishing. All rights reserved.)

It is clear that the inability of FIR experiments to detect charge disproportionation in $\kappa$-(ET)$_2$Cu[N(CN)$_2$]Cl may represent a problem for an interpretation along the lines of figure 23. However, as pointed out in ref. [137] the lower limit of charge disproportionation of $0.005\,e$ obtained from the FIR experiments may still be compatible with the size of the observed dielectric anomaly if considering the results in the charge ordered (TMTTF)$_2X$ salts (cf. section 4.1). However, this notion is based on some rough estimates and a more thorough, theoretical treatment would be desirable. Moreover, in ref. [139] it was considered that there may be indeed a CO transition in $\kappa$-(ET)$_2$Cu[N(CN)$_2$]Cl, which in some nominally pure single crystals is smeared out and shifted to higher temperatures due to internal inhomogeneities. Indeed in ref. [137] a considerable sample-to-sample variation of the conductivity anomaly and the dielectric response was found by comparing results in eight single crystals of which three even showed relaxor-type instead of long-range ferroelectricity. Clearly more experimental efforts are necessary to finally resolve the interesting problem of the dielectric properties of $\kappa$-(ET)$_2$Cu[N(CN)$_2$]Cl.

*4.2.4 Other systems*

Finally, we want to mention here that indications for relaxor ferroelectricity were also found for $\lambda$-(BEDT-TSF)$_2$FeCl$_4$ (BEDT-TSF stands for bisethylenedithiotetraselenafulvalene) based on microwave measurements, which will not be treated in detail here [120]. Later on, strong magnetodielectric effects were detected in this system and a CO-induced polarization model was proposed to explain its dielectric properties [23].

Another 2-dimensional system revealing interesting dielectric properties is $\Theta$-(ET)$_2$CsZn(SCN)$_4$. In [148] clear evidence for a relaxational process in this system was found by dielectric measurements in the kHz to MHz range. It could be well described by the HN function, eq. (1). Moreover, a marked bias dependence of the dielectric properties was found. In [148] these results were interpreted in terms of collective excitations of a CDW.

## 5. Summary and Conclusions

The preceding sections have shown that organic charge-transfer salts exhibit a rich "zoo" of dielectric phenomena, allowing to draw conclusions on such interesting properties as electronic ferroelectricity, relaxor ferroelectricity and multiferroicity. Dipolar relaxation, dc charge transport, ac conductivity arising from hopping conductivity, dielectric anomalies due to short- and long-range ferroelectricity and Maxwell-Wagner relaxations all can be found in these materials. As these processes can partly occur simultaneously in the same material, care should be taken in the interpretation of the dielectric data and measurements should be made in a frequency and temperature range that is as broad as possible. Moreover, considerable sample-to-sample variations may pose problems in the interpretation of data.

In this review, we have concentrated on two classes of organic charge-transfer salts, showing especially interesting dielectric phenomena. The one-dimensional systems treated in section 4.1 have proven to be ideally suited for studying the two fundamentally different mechanisms of ferroelectricity, namely the generation of polar order by ionic or electronic degrees of freedom. Especially the investigation of systems like TTF-CA, where both mechanisms compete [6], to us seems a promising task for further research.

The two-dimensional $(ET)_2X$ systems stand out by the large variation of properties that partly can be ascribed to the different arrangements of the molecules within the ET layers. However, even for the two $\kappa$-type systems, $\kappa$-$(ET)_2Cu_2(CN)_3$ and $\kappa$-$(ET)_2Cu[N(CN)_2]Cl$, markedly different dielectric behaviour is found [139]. The first compound is characterized by dipolar and magnetic degrees of freedom that, however, both do not develop long-range order [8]. In contrast, the latter system was found to be multiferroic, showing both dipolar and magnetic order and an especially close coupling of the magnetic and dipolar degrees of freedom, for which a new type of electric-dipole driven multiferroic mechanism was proposed [7]. Frustration effects arising from the triangular lattices within the ET planes, which exhibit different anisotropies, may explain these different low-temperature phases of both materials [139] but the details are far from being understood. In fact, even the nature of the dipolar degrees of freedom in these systems is not yet clarified, especially concerning the role of CO, for which no clear evidence was found until now [126]. Interestingly, in $\beta'$-$(ET)_2ICl_2$ not exhibiting the triangular arrangement of the other systems, an absence of long-range polar order was detected, too, while magnetic order is known to occur in this system [9]. Another intriguing example is $\alpha$-$(ET)_2I_3$ where a complex charge-order pattern within the ET planes seems to be the reason for the occurrence of relaxor ferroelectricity at temperatures significantly below $T_{CO}$ [10]. Overall, there is a clear need for further thorough investigations of the dielectric properties of organic charge-transfer salts, in order to arrive at a concise and systematic picture of the mechanisms governing their intriguingly rich properties.

**Acknowledgments**

We thank M. Lang and J. Müller from the Goethe-University Frankfurt for the very fruitful cooperation within the field of organic charge-transfer salts. We are grateful to J. Schlueter and D. Schweitzer for providing the samples for our dielectric measurements. We thank the authors and publishers of refs. [1,4,6,8,9,22,24,79,81,136,138,146] for their kind permission for using figures from their works. This work was supported by the Deutsche Forschungsgemeinschaft through the Transregional Collaborative Research Center TRR 80.

**References**

[1]   Nad F and Monceau P 2006 *J. Phys. Soc. Japan* **75** 051005
[2]   Monceau P, Nad F Ya and Brazovskii S 2001 *Phys. Rev. Lett.* **86** 4080

[3]   Nad F, Monceau P, Kaboub L and Fabre J M 2006 *Europhys. Lett.* **73** 567
[4]   Starešinić D, Biljaković K, Lunkenheimer P and Loidl A 2006 *Solid State Commun.* **137** 241
[5]   Kagawa F, Horiuchi S, Tokunaga M, Fujioka J and Tokura Y 2010 *Nature Phys.* **6** 169
[6]   Kobayashi K, Horiuchi S, Kumai R, Kagawa F, Murakami Y and Y Tokura 2012 *Phys. Rev. Lett.* **108** 237601
[7]   Lunkenheimer P, Müller, J, Krohns S, Schrettle F, Loidl A, Hartmann B, Rommel R, de Souza M, Hotta C, Schlueter J A and Lang M 2012 *Nature Mater.* **11** 755
[8]   Abdel-Jawad M, Terasaki I, Sasaki T, Yoneyama N, Kobayashi N, Uesu Y and Hotta C 2010 *Phys. Rev. B* **82** 125119
[9]   Iguchi S, Sasaki S, Yoneyama N, Taniguchi H, Nishizaki T and Sasaki T 2013 *Phys. Rev. B* **87** 075107
[10]  Lunkenheimer P, Hartmann B, Lang M, Müller J, Schweitzer D, Krohns S and Loidl A, arXiv:1407.0339
[11]  Ivek T, Korin-Hamzić B, Milat O, Tomić S, Clauss C, Drichko N, Schweitzer D and Dressel M, 2010 *Phys. Rev. Lett.* **104** 206406
[12]  Takahashi T, Nogami Y and Yakushi K 2006 *J. Phys. Soc. Jpn.* **75** 051008
[13]  Mercone S, Wahl A, Pautrat A, Pollet M and Simon C 2004 *Phys. Rev. B* **69** 174433
[14]  Park T *et al.* 2005 *Phys. Rev. Lett.* **94** 017002
[15]  Krohns S , Lunkenheimer P, Kant Ch, Pronin A V, Brom H B, Nugroho A A, Diantoro M and Loidl A 2009 *Appl. Phys. Lett.* **94** 122903
[16]  Ishihara S 2014 *J. Phys.: Condens. Matter* **26** 493201
[17]  Hill N A 2000 *J. Chem. Phys. B* **104** 6694
[18]  Fiebig M 2005 *J. Phys. D: Appl. Phys.* **38** R123
[19]  Eerenstein W, Mathur N D and Scott J F 2006 *Nature* **442** 759
[20]  Cheong S-W and Mostovoy M 2007 *Nature Materials* **6** 13
[21]  Van den Brink J and Khomskii D I 2008 *J. Phys. Condens. Matter* **20** 434217
[22]  Tomić S, Pinterić M, Ivek T, Sedlmeier K, Beyer R, Wu D, Schlueter J A, Schweitzer D and Dressel M 2013 *J. Phys.: Condens. Matter* **25** 436004
[23]  Negishi E, Kuwabara T, Komiyama S, Watanabe M, Noda Y, Mori T, Matsui H and Toyota N 2005 *Phys. Rev. B* **71** 012416
[24]  Kodama K, Kimata M, Takahide Y, Kurita N, Harada A, Satsukawa H, Terashima T, Uji S, Yamamoto K and Yakushi K 2012 *J. Phys. Soc. Jpn.* **81** 044703
[25]  Kremer F and Schönhals A 2003 In: *Broadband dielectric spectroscopy* (eds. Kremer F and Schönhals A) p. 35 (Springer: Berlin).
[26]  Jonscher A K 1983 *Dielectric Relaxation in Solids* (London: Chelsea Dielectrics Press)
[27]  Böhmer R, Maglione M, Lunkenheimer P and Loidl A 1989 *J. Appl. Phys.* **65** 901
[28]  Scheffler M, Fella C and Dressel M 2012 *J. Physics: Conf. Series* **400** 052031
[29]  Klein O, Donovan S, Dressel M and Grüner G 1993 *Int. J. Infrared and Millimeter Waves* **14** 2423
[30]  Scheffler M and Dressel M 2005 *Rev. Sci. Instrum.* **76** 074702
[31]  Schneider U, Lunkenheimer P, Pimenov A, Brand R and Loidl A 2011 *Ferroelectrics* **249** 89
[32]  Gorshunov B, Volkov A A, Spektor I, Prokhorov A M, Mukhin A, Dressel M, Uchida S and Loidl A 2005 *Int. J. Infrared and Millimeter Waves* **26**, 1217
[33]  Kremer F and Schönhals A (eds.) 2003 *Broadband dielectric spectroscopy* (Springer: Berlin).
[34]  Lunkenheimer P, Schneider U, Brand R and Loidl A 2000 *Contemp. Phys.* **41** 15
[35]  Fehst I, Paasch M, Hutton S L, Braune M , Böhmer R, Loidl A, Dörffel M, Narz Th, Haussühl S and Mcintyre G J 1993 *Ferroelectrics* **138** 1
[36]  Hutton S L, Fehst I, Böhmer R, Braune M, Mertz B, Lunkenheimer P and Loidl A 1991 *Phys. Rev. Lett.* **66** 1990
[37]  Debye P 1929 *Polare Molekeln* (Hirzel: Leipzig).
[38]  Böttcher C J F and Bordewijk P 1973 *Theory of electric polarization Vol II* (Elsevier: Amsterdam)
[39]  Havriliak S and Negami S 1966 *J. Polymer Sci. C* **14**, 99
[40]  Cole K S and Cole R H 1941 *J. Chem. Phys.* **9**, 341
[41]  Davidson D W and Cole R H 1950 *J. Chem. Phys.* **18**, 1417
[42]  Davidson D W and Cole R H 1951 *J. Chem. Phys.* **19** 1484

[43] Sillescu H 1999 *J. Non-Cryst. Solids* **243** 81
[44] Ediger M D 2000 *Annu. Rev. Phys. Chem.* **51** 99
[45] Vogel H 1921 *Phys. Z.* **22** 645
[46] Fulcher G S 1925 *J. Am. Ceram. Soc.* **8** 339
[47] Tammann G and Hesse W 1926 *Z. Anorg. Allg. Chem.* **156** 245
[48] Viehland D, Jang S J, Cross L E and Wuttig M 1990 *J Appl Phys* **68** 2916
[49] Levstik A, Kutnjak Z, Filipiĉ C and Pirc R 1998 *Phys. Rev. B* **57** 11204
[50] Glazounov A E and Tagantsev A K 1998 *Appl. Phys. Lett.* **73** 856
[51] Hiraki K and Kanoda K 1998 *Phys. Rev. Lett.* **80** 4737
[52] Maki K, Dóra B, Kartsovnik M, Virosztek A, Korin-Hamzić B, and Basletić M 2003 *Phys. Rev. Lett.* **90** 256402
[53] Jérome D 2004 *Chem. Rev.* **104** 5565
[54] Grüner G 1988 *Rev. Mod. Phys.* **60** 1129
[55] Littlewood P B 1987 *Phys. Rev. B* **36** 3108
[56] Pike G E 1972 *Phys. Rev. B* **6** 1572
[57] Lunkenheimer P, Loidl A, Ottermann C R and K Bange 1991 *Phys. Rev. B* **44** 5927
[58] Lunkenheimer P, Resch M, Loidl A and Hidaka Y1992 *Phys. Rev. Lett.* **69** 498
[59] Seeger A, Lunkenheimer P, Hemberger J, Mukhin A A, Ivanov V Yu, Balbashov A M and Loidl A 1999 *J. Phys.: Condens. Matter* **11** 3273
[60] Sichelschmidt J *et al*. 2001 *Eur. Phys. J. B* **20** 7
[61] Zhang L and Tang Z J 2004 *Phys. Rev. B* **70** 174306
[62] Renner B, Lunkenheimer P, Schetter M, Loidl A, Reller A and Ebbinghaus S G 2004 *J Appl Phys* **96** 4400
[63] Sano K, Sasaki T, Yoneyama N and Kobayashi N 2010 *Phys. Rev. Lett.* **104** 217003
[64] Köhler B, Rose E, Dumm M, Untereiner G and Dressel M 2011 *Phys. Rev. B* **84** 035124
[65] Pinterić M *et al*. 2014 *Phys. Rev. B* **90** 195139
[66] Long A R 1982 *Adv. Phys.* **31** 553
[67] Elliott S R 1987 *Adv. Phys.* **36** 135
[68] van Staveren M P J, Brom H B and de Jongh L 1991 *Phys. Reports* **208** 1
[69] Dyre J C and Schrøder T B 2000 *Rev. Mod. Phys.* **72** 873
[70] Mott N F and Davies E A 1979 *Electronic Processes in Non-Crystalline Materials* (Oxford: Oxford University Press)
[71] Shante V K, Varma C M and Bloch A N 1973 *Phys. Rev. B* **8** 4885
[72] Efros A L and Shklovskii B I 1975 *J. Phys. C* **8** L49
[73] Devonshire A F 1954 *Adv. Phys.* **3** 85
[74] Cochran W 1960 *Adv. Phys.* **9** 387
[75] Jona F and Shirane G 1962 *Ferroelectric Crystals* (London: Pergamon Press)
[76] Blinc R and Zeks B 1974 *Soft Modes in Ferroelectrics and Antiferroelectrics* (Amsterdam: North Holland Publishing)
[77] Lines M E and Glass A M 1977 *Principles and Applications of Ferroelectrics and Related Materials* (Clarendon: Oxford)
[78] Cowley R A 1962 *Phys. Rev. Lett.* **9** 159
[79] Johnson C J 1965 *Appl. Phys. Lett.* **7** 221
[80] Yamada Y, Fujii Y and Hatta I 1968 *J. Phys. Soc. Japan* **24** 1053
[81] Gesi K J. 1970 *Phys. Soc. Japan* **28** 1365
[82] Smolenski G A *et al*. 1960 Sov. *Phys. Solid State* **2** 2584
[83] Höchli U T, Knorr K and Loidl A 1990 *Adv. Phys.* **39** 405
[84] Cross L E 1987 *Ferroelectrics* **76** 241
[85] Samara G A 2003 J. *Phys. Condens. Matt.* **15** R367
[86] Ang C, Yu Z, Lunkenheimer P, Hemberger J, and Loidl A 1999 *Phys. Rev. B* **59** 6670
[87] Viehland D, Jang S J, Cross L E and Wuttig M 1992 *Phys. Rev. B* **46** 8003
[88] Kimura T, Goto T, Shintal H, Ishizaka K, Arima T and Tokura Y 2003 *Nature* **426** 55
[89] Mostovoy M 2006 *Phys. Rev. Lett.* **96** 067601
[90] Verwey E 1839 *Nature* **144** 327

[91]   J Maxwell J C 1873 *A Treatise on Electricity and Magnetism* (Clarendon Press: Oxford)
[92]   Wagner K W 1914 *Arch. Electrotech.* **2** 371
[93]   Sinclair D C, Adams T B, Morrison F D, and West A R 2002 *Appl. Phys. Lett.* **80** 2153
[94]   Lunkenheimer P, Bobnar V, Pronin A V, Ritus A I, Volkov A A and Loidl A 2002 *Phys. Rev. B* **66** 052105
[95]   Lunkenheimer P, Fichtl R, Ebbinghaus S G and Loidl A 2004 *Phys. Rev. B* **70** 172102
[96]   Lunkenheimer P, Krohns S, Riegg S, Ebbinghaus S G, Reller A and Loidl A 2010 *Eur Phys J Special Topics* **180** 61
[97]   Bobnar V, Lunkenheimer P, Paraskevopoulos M and Loidl A 2002 *Phys. Rev. B* **65**, 184403
[98]   Krohns S, Lunkenheimer P, Ebbinghaus S G and Loidl A 2007 *Appl. Phys. Lett.* **91** 022910
[99]   Pouget J-P 2012 *Physica B* **407** 1762
[100]  Nad F, Monceau P and Fabre J M 1998 *Eur. Phys. J. B* **3** 301
[101]  Nad F, Monceau P, Carcel C and Fabre J M 2000 *Phys. Rev. B* **62** 1753
[102]  Nad F, Monceau P, Carcel C and Fabre J M 2000 *J. Phys.: Condens. Matter* **12** L435
[103]  Nagasawa M, F Nad F, Monceau P and Fabre J M 2005 *Solid State Commun.* **136** 262
[104]  Barzovskii S, Monceau P and Nad F 2015 *Physica B* **460** 76
[105]  Chow D S, Zamborszky F, Alavi B, Tantillo D J, Baur A, Merlic C A and Brown S E 2000 *Phys. Rev. Lett.* **85** 1698
[106]  Zamborszky F, Yu W, Raas W, Brown S E, Alavi B, Merlic C A and Baur A 2002 *Phys. Rev. B* **66** 081103
[107]  Zorina L, Simonov S, Meziere C, Canadell E, Suh S, Brown S E, Foury-Leylekian P, Fertev P, Pouget J-P and Batail P 2009 *J. Mater. Chem.* **19** 6980
[108]  Riera J and Poilblanc D 2001 *Phys. Rev. B* 63 241102
[109]  Yu W, Zhang F, Zamborszky F, Alavi B, Baur A, Merlic C A and Brown S E 2004 *Phys. Rev. B* **70** 121101
[110]  Pouget J-P, Foury-Leylekian P, Alemany P and Canadell E 2012 *Phys. Status Solidi B* **249** 937
[111]  Yu W, Zamborszky F, Alavi B, Baur A, Merlic C A and Brown S E 2004 *J. Phys. IV France* **114**, 35
[112]  Giovannetti G, Nourafkan R, Kotliar G and Capone M 2015 *Phys. Rev. B* **91** 125130
[113]  Yoshimi K, Seo H, Ishibashi S and Brown S E 2012 *Phys. Rev. Lett.* **108**, 096402
[114]  Macedo P B, Moynihan C T, and Bose R 1972 *Phys. Chem. Glasses* **13** 171
[115]  Tokura Y, Koshihara S, Iwasa Y, Okamoto H, Komatsu T, Koda T, Iwasawa N and Saito G 1989 *Phys. Rev. Lett.* **63**, 2405
[116]  Horiuchi S, Kobayashi K, Kumai R and Ishibashi S 2014 *Chem. Lett.* **43** 26
[117]  Torrance J B, Vazquez J E, Mayerle J J and Lee V Y 1981 *Phys. Rev. Lett.* **46** 253
[118]  Giovannetti G, Kumar S, Stroppa A, van den Brink J and Picozzi S 2009 *Phys. Rev. Lett.* **103** 266401
[119]  Okamoto H, Mitani T, Tokura Y, Koshihara S, Komatsu T, Iwasa Y, Koda T and Saito G 1991 *Phys. Rev. B* **43** 8224
[120]  Matsui H, Tsuchiya H, Suzuki T, Negishi E and Toyota N 2003 *Phys. Rev. B* **68** 155105
[121]  Sasaki T, Yoneyama N, Nakamura Y, Kobayashi N, Ikemoto Y, Moriwaki T and Kimura H 2008 *Phys. Rev. Lett.* **101** 206403
[122]  Shimizu Y, Miyagawa K, Kanoda K, Maesato M and Saito G 2003 *Phys. Rev. Lett.* **91** 107001
[123]  Naka M and Ishihara S 2010 *J. Soc. Phys. Jpn.* **79**, 063707
[124]  Hotta C 2010 *Phys. Rev. B* **82** 241104(R)
[125]  Gomi H, Ikenaga M, Hiragi Y, Segawa D, Takahashi A, Inagaki T J and Aihara M 2013 Phys. Rev. B **87** 195126
[126]  K Sedlmeier, Elsässer S, Neubauer D, Beyer R, Wu D, Ivek T, Tomić S, Schlueter J A and Dressel M 2012 *Phys. Rev. B* **86** 245103
[127]  Yoneyama N, Miyazaki A, Enoki T and Saito G 1999 *Bull. Chem. Soc. Jpn.* **72** 639
[128]  Yamamoto K, Iwai S, Boyko S, Kashiwazaki A, Hiramatsu F, Okabe C, Nishi N and Yakushi K 2008 *J. Soc. Phys. Jpn.* **77** 074709
[129]  Yamamoto K, Kowalska A A and Yakushi K 2010 *Appl. Phys. Lett.* **96** 122901
[130]  Bloembergen N and Pershan P S 1962 *Phys. Rev.* **128** 606
[131]  Williams D J 1984 *Angew. Chem. Int. Ed. Engl.* **23** 690

[132] Bender K, Hennig I, Schweitzer D, Dietz K, Endres H and H J Keller 1984 *Mol. Cryst. Liq. Cryst.* **108** 359
[133] Fortune N A, Murata K, Ishibashi M, Tokumoto M, Kinoshita N and Anzai H 1991 *Solid State Commun.* **79** 265
[134] Bender K, Dietz K, Endres H, Helberg H W, Hennig I, Keller H J, Schafer H W and Schweitzer D 1984 *Mol. Cryst. Liq. Cryst.* **107** 45
[135] Dressel M, Grüner G, Pouget J P, Breining A and Schweitzer D 1994 *J. Phys. I France* **4** 579
[136] Ivek T, Korin-Hamzić B, Milat O, Tomić S, Clauss C, Drichko N, Schweitzer D and Dressel M 2011 *Phys. Rev. B* **83** 165128
[137] Lang M, Lunkenheimer P, Müller J, Loidl A, Hartmann B, Hoang N H, Gati E, Schubert H and Schlueter J A 2014 *IEEE Trans. Mag.* **50**, 2700107
[138] Pinterić M, Miljak M, Biškup N, Milat O, Aviani I, Tomić S, Schweitzer D, Strunz W and Heinen I 1999 *Eur. Phys. J. B* **11** 217
[139] Pinterić M, Ivek T, Čulo M, Milat O, Basletić M, Korin-Hamzić B, Tafra E, Hamzić A, Dressel M, Tomić S 2015 *Physica B* **460** 202
[140] Miyagawa K, Kawamoto A, Nakazawa Y and Kanoda K 1995 *Phys. Rev. Lett.* **75**, 1174
[141] Smith D F, De Soto S M, Slichter C P, Schlueter K A, Kini A M and Daugherty R G 2003 *Phys. Rev. B* **68** 024512
[142] Dressel M, Grüner G, Carlson K D, Wang H H and Williams J M 1995 *Synth. Met.* **70**, 927
[143] Ito H, Ishiguro T, Kubota M and Saito G 1996 *J. Phys. Soc. Japan* **65**, 2987
[144] Williams J M, Kini A M, Wang H H, Carlson K D, Geiser U, Montgomery L K, Pyrka G J, Watkins D M and Kommers J M 1990 *Inorg. Chem.* **29** 3272
[145] Elsässer S, Wu D, Dressel M and Schlueter J A 2012 *Phys. Rev. B* **86** 155150
[146] Takahide Y, Kimata M, Kodama K, Terashima T, Uji S, Kobayashi M and Yamamoto H M 2011 *Phys. Rev B* **84**, 035129
[147] Sergienko I A and Dagotto, E. 2006 *Phys. Rev. B* **73**, 094434
[148] Inagaki K, Terasaki I, Mori H and Mori T 2004 *J. Phys. Soc. Japan* **73**, 3364